\documentstyle[]{article}  

\begin{document}
\newcommand {\bea}{\vspace{-.00in}\begin{eqnarray}}   
\newcommand {\eea}{\vspace{-.00in}\end{eqnarray}}
\newcommand {\beaa}{\vspace{-.00in}\begin{eqnarray*}}   
\newcommand {\eeaa}{\vspace{-.00in}\end{eqnarray*}}
\newcommand {\be}{\vspace{-.00in}\begin{equation}}   
\newcommand {\ee}{\vspace{-.00in}\end{equation}}
\newcommand {\eps} {\epsilon}
\newcommand {\Del} {\Delta}
\newcommand {\bm}  {\boldmath}
\newcommand {\del} {\delta}
\newcommand {\sig} {\sigma}
\newcommand {\nn} {\nonumber}
\newcommand  {\th} {P_H}
\newcommand  {\tl} {P_L}
\newcommand {\da} {\dagger}
\newcommand  {\lam} {\lambda}
\newcommand  {\bp}  {|{\bf p}|}
\newcommand  {\pr} {\prime}
\newcommand  {\kap} {\kappa}
\newcommand   {\hp}  {\hat{{\bf p}}}
\newcommand   {\lr}  {\leftrightarrow}
\newcommand  {\Lam} {\Lambda}
\newcommand  {\gam} {\gamma}
\newcommand  {\ren} {{\lam_{_{R}}}}
\newcommand  {\veps} {\varepsilon}
\newcommand   {\pp}   {({\bf p}-{{\bf p}^\pr})^2}
\newcommand   {\easy}  {\del M_{_{1}}^2}
\newcommand    {\hard} {\del M_{_{2}}^2}
\newcommand    {\p} {\Omega_p}
\newcommand    {\ppr} {\Omega_{p^\prime}}
\newcommand    {\k} {\Omega_k}
\newcommand    {\kpr} {\Omega_{k^\prime}}
\newcommand    {\pig} {\left(\int_0^\pi \frac{d\omega \sin^2\omega}
{1+\cos\omega} f_{2,1}(\omega) f_{n,1}(\omega)\right)^2}
 \newcommand    {\tp} {{\tilde{\phi}}}
 \newcommand    {\tP} {{\tilde{\Phi}}}
\centerline{\large {\bf The Lamb Shift in a Light Front Hamiltonian Approach
}}
\vskip.1in
\centerline{Billy D. Jones and Robert J. Perry}
\centerline{  Department of Physics, The Ohio State University, Columbus, Ohio 43210-1106}
\centerline{\today}
\vskip.1in
Light-front Hamiltonian methods are being developed
to attack bound-state problems in QCD. In this paper we advance the state of the art 
for these methods by computing the well-known Lamb shift in hydrogen starting 
from first principles of QED. There are obvious but significant qualitative differences
between QED and QCD. In this paper,  we discuss the
 similarities that may survive in a non-perturbative
QCD calculation in the context of a precision non-perturbative QED calculation. Central
to the discussion are how a constituent picture arises in a gauge field theory, 
 how bound-state energy scales 
emerge to guide the renormalization procedure, and how
 rotational invariance emerges in a light-front
calculation.
\vskip.2in
\noindent
PACS number(s): 11.10.Ef, 12.20.Ds
\vskip.3in
\centerline{{\bf I. INTRODUCTION}}
\vskip.2in

Why is the calculation of the Lamb shift in hydrogen, which at the level of
detail found in this paper was largely
completed by Bethe in 1947 \cite{bethe}, of any real
interest today?  While completing such a calculation using new techniques may be
very interesting for formal and academic reasons, our primary motivation is to
lay groundwork for precision bound-state calculations in QCD.  The Lamb shift
provides an excellent pedagogical tool for illustrating light-front Hamiltonian
techniques, which are not widely known; but more importantly it presents three
of the central dynamical and computational problems that we must face to make
these techniques useful for solving QCD: How does a constituent picture emerge in
a gauge field theory?  How do bound-state energy scales emerge
non-perturbatively?  How does rotational symmetry emerge in a non-perturbative
light-front calculation?  

These questions can be answered in detail in QED.  The answers clearly change in
QCD, and we point out several places where this is clear, but we hope that much
of the computational framework successfully employed in QED will survive.

In order to formulate these questions in a more precise fashion, we first
outline the general computational strategy we employ.  First, we  use
the renormalization group
to produce a regulated effective Hamiltonian $H_\lambda$, where
$\lambda$ is a cutoff and renormalization is required to remove cutoff
dependence from all physical quantities.  At this point we have a regulated
Hamiltonian that contains all interactions found in the canonical Hamiltonian,
a finite number of new relevant and marginal operators (each of which contains
a function of longitudinal momenta because longitudinal locality is not
maintained in light-front field theory), and an infinite number of irrelevant
operators as would occur in any cutoff theory.  This complicated Hamiltonian
cannot be directly diagonalized, and since we want to solve bound-state
problems we cannot solve it using perturbation theory.  The second step is to
approximate the full Hamiltonian, using

\bea
 &&H_\lam={\cal H}_o + \left(H_\lam-{\cal H}_o\right)~\equiv~{\cal H}_o+{\cal V}~,
\label{eq:1cv}
 \eea

\noindent where ${\cal H}_o$ is an approximation that can be solved
non-perturbatively and ${\cal V}$ is treated in bound-state perturbation theory (BSPT). 
The test of ${\cal H}_o$ is whether BSPT converges or not.

We can now reformulate the questions above.  Is there a scale $\lambda$ at which
${\cal H}_o$ does not require particle emission and absorption?  What are the few-body
interactions in ${\cal H}_o$ that generate the correct non-perturbative bound-state
energy scales?  Is there a few-body realization of rotational invariance; and
if not, how does rotational symmetry emerge in BSPT? 
We should emphasize that for our purposes we are primarily interested in
answering these questions for low-lying bound-states, and refinements may be
essential to discuss highly excited states or bound-state scattering.

It is essential that $\lambda$, which governs the degree to which states are resolved,
 be adjusted to obtain a constituent approximation.  If $\lambda$ is
kept large with respect to all mass scales in the problem, arbitrarily large
numbers of constituents are required in the states because constituent
substructure is resolved.  A constituent picture can emerge if high free-energy
states couple perturbatively to the low free-energy states that dominate the
low-lying bound-states.  In this case the cutoff can be lowered until it
approaches the non-perturbative bound-state energy scale and perturbative
renormalization may be employed to approximate the effective Hamiltonian.  In
QED we note that the range into which the cutoff must be lowered is

\begin{equation}
m\alpha^2  \ll \tilde{\lambda} \ll m\alpha \;\;,
\label{eq:2cv}
\end{equation}

\noindent where $\tilde{\lambda} = \lambda - m_p - m_e$ as will be explained
later, and $m$ is the reduced mass of hydrogen. 
 If the cutoff is lowered to this range, hydrogen bound-states are well
approximated using proton-electron states and including photons and pairs
perturbatively.

It is an oversimplification to say the constituent picture emerges because the
QED coupling constant is very small.  Photons are massless, and regardless of how
small $\alpha$ is, one must in principle use nearly degenerate bound-state
perturbation theory that includes extremely low energy photons
non-perturbatively.  This is not required in practice, because the Coulomb interaction which
sets the important energy scales for the problem produces neutral bound-states
from which long wavelength photons effectively decouple.  Because of this, even
though arbitrarily small energy denominators are encountered in BSPT
 due to mixing of electron-proton bound-states and states
including extra photons, BSPT can converge because
emission and absorption matrix elements vanish sufficiently rapidly.

The well-known answer to the second question above is the two-body Coulomb
interaction sets the non-perturbative energy and momentum scales appropriate
for  QED.  We have already used the results of the Bohr scaling analysis that
reveals the bound-state momenta scale as $p \sim m\alpha$ and the energy
scales as $E \sim m\alpha^2 $.  As a result the dominant photon momenta are
also of order $m\alpha $, and the corresponding photon energies are of
order $m\alpha $.  This is what makes it possible to use renormalization to
replace photons with effective interactions.  The dominant photon energy scale
is much greater than the bound-state energy scale, so that $\lambda$ can be
perturbatively lowered into the window in Eq.~(\ref{eq:2cv}) and photons are not
required in the state to leading order.  A similar analysis in QCD
will reveal qualitatively different results.  If a constituent picture emerges,
the dominant interaction will be confining and the dominant gluon energy scale
will be directly affected by confinement.  A confining interaction
automatically generates a mass gap for gluon production.

Finally we discuss rotational invariance in a light-front approach.  In
light-front field theory, boost invariance is kinematic, but rotations about
transverse axes involve interactions.  Thus rotational invariance is not
manifest and all cutoffs violate rotational invariance in light-front field
theories. In QED it is easy to see how counterterms in $H_\lambda$ arise during
renormalization that repair this symmetry perturbatively; however, the issue of
non-perturbative rotational symmetry is potentially much more complicated.  We
first discuss leading order BSPT and then turn
to higher orders.

To leading order in a constituent picture we require a few-body realization of
rotational symmetry.  This is simple in non-relativistic systems, because
Galilean rotations and boosts are both kinematic.  In QED the constitutuent
momenta in all low-lying bound-states are small, so a non-relativistic reduction
can be used to derive ${\cal H}_o$.  Therefore to leading order in QED we can employ a
non-relativistic realization of rotational invariance.  This type of approach
can be tried in QCD, but it is not essential that it work because alternative
few-body realizations of the full set of Lorentz symmetries exist.

At higher orders in BSPT rotational invariance will
not be maintained unless corrections are regrouped.  We have computed hyperfine
structure and shown that terms from first-order and second-order BSPT
 are required to obtain angular momentum multiplets \cite{billy}.  The
guiding principle in this and all higher order calculations is to expand not in
powers of ${\cal V}$, but in powers of $\alpha$ and $\log(\alpha)$.  ${\cal H}_o$ should
provide the leading term in this expansion for BSPT
to be well-behaved, and subsequent terms should emerge from finite orders of
BSPT after appropriate regrouping.  Powers of
$\alpha$ appear through explicit dependence of interactions on $\alpha$, and
through the dependence of leading order eigenvalues and eigenstates on $\alpha$
introduced by interactions in ${\cal H}_o$.  This second source of dependence can be
estimated using the fact that momenta scale as $m\alpha$ in the bound-state
wave functions.  Of more interest for this paper is the appearance of
$\log(\alpha)$, which is signaled by a divergence in unregulated bound-state
perturbation theory.  As has long been appreciated, such logarithms appear
when the number of scales contributing to a correction diverges.

The existence of a small parameter simplifies the non-perturbative calculation
of bound-state observables considerably, and it has been suggested that a
similar expansion be employed to guide light-front QCD calculations even if it
requires the introduction of masses that violate rotational invariance away from
the critical value of the coupling \cite{longpaper}.  We do not detail this proposal, but a
thorough understanding of such expansions in QED is almost certainly necessary
before one has any hope of using this approach for QCD.

We proceed with a description of our Lamb shift calculation.  In hydrogen there
is a small amplitude for a bound electron to emit and re-absorb a photon, which
leads to a small shift in the binding energy.  This is the dominant source of
the Lamb shift, and the only part of this shift we compute in this paper.
This requires electron self-energy renormalization, but 
removal of all
the bare cutoff, $\tilde{\Lam}$, dependence requires a complete 4th order calculation,
which is beyond the scope of 
this paper. We work with a finite bare cutoff: $\tilde{\Lam}=m\sqrt{2}$, and show that our
results are independent of the effective cutoff, $\tilde{\lam}$.

The energy scale for the electron binding energy is $m\alpha^2$, while the
scale for photons that couple to the bound-states is $m\alpha$.  This energy
gap makes the theory amenable to the use of effective Hamiltonian techniques. 
For simplicity, we use a Bloch transformation \cite{bloch}
in this paper to remove the high energy scale
({\it i.e.}, $m\alpha $) from the states, and an effective Hamiltonian is
derived which acts in the low energy space alone.  This effective Hamiltonian
is treated in BSPT, as outlined above.  The
difference between the $2 S_{1 \over 2}$ and the $2 P_{1 \over 2}$ energy
levels, which are degenerate to lowest order, is calculated.

We divide the calculation into two parts, low and high energy intermediate
photon contributions.  The low energy photons satisfy $|{\bf k}| <
\tilde{\lambda}$, while the high energy intermediate photons satisfy
$\tilde{\lambda} < |{\bf k}| < m$.  $\tilde{\lambda}$ is the effective cutoff for
the theory, which is chosen to lie in the range given in Eq.~(\ref{eq:2cv}). This choice
lies between the two dominant energy scales in the problem and allows us to
avoid near degeneracy problems.  When an actual number is required we use

\begin{equation}
\tilde{\lambda} = \alpha\sqrt{\alpha}\; m \sim 6 \times 10^{-4}\; m \;.
\label{eq:32415}\end{equation}

\noindent Note that the spectrum of the exact effective Hamiltonian is
independent of $\tilde{\lambda}$, but our approximations introduce
$\tilde{\lambda}$-dependence.  The range for $\tilde{\lambda}$ is chosen so
that the errors appear at a higher order in $\alpha$ than we compute.

One further introductory comment, the high photon energy ($\tilde{\lam}<|{\bf
k}| < m$)  part of the shift is further divided into two regions,
$\tilde{\lam}<|{\bf k}| < b$ and $b<|{\bf k}| < m$, where $b$ is an arbitrary
parameter chosen in the range $m \alpha \ll b \ll m$. This  simplifies the
calculation with appropriate approximations being used in the respective
regions. The result must obviously be independent of this arbitrary division
point $b$, and is, unless ``non-matching" approximations are used in the
respective regions.

We now outline the paper. In \S II we discuss the theoretical framework of this
light-front Hamiltonian approach, and in \S III we proceed to  discuss  the
origin of the Coulomb interaction in this framework. \S IV contains the heart of
the Lamb shift calculation. In the final section, \S V, we summarize and
discuss our results.

\vskip.2in
\noindent
\centerline{{\bf II. THEORETICAL FRAMEWORK}}
\vskip.2in

In this paper, the proton will be treated as a point
particle. 
The Lagrangian for the electron, proton, and photon system is ($e > 0$)
\bea
{\cal L}&=&-\frac{1}{4}F_{\mu \nu} F^{\mu \nu} +
\overline{\psi}_e(i \not\! {\partial } +e \not\!\! {A }- 
 m_e )\psi_e+
 \overline{\psi}_p(i \not\! {\partial } -e \not\!\! {A }- 
 m_p )\psi_p~.
\eea
The reduced mass of
the system is defined in the standard way
\bea
m&=&\frac{m_e m_p}{m_e+m_p}=m_e \left(1-m_e/m_p+{\cal O}(1/m_p^2)\right)
\;.\label{eq:reduced}
\eea
Note that in this paper we take the limit $m_p \longrightarrow \infty$
because we are only interested in the dominant part of the Lamb shift.
The Lagrangian leads to the following canonical Hamiltonian in the light-cone gauge, $A^+ =0$,
\bea
&&H=
 \int d^2 x^{\perp}dx^- ~{\cal H}~~,
 \label{eq:can1}\\
 \nn\\
 &&{\cal H}=
 \frac{1}{2} \left(\partial^i A^j\right)^2+
 \xi_e^\dagger\left(i\hspace{.1mm}\sigma^i\partial^i+
 e\hspace{.1mm}\sig^iA^i-i\hspace{.1mm} m_e\right)
 \frac{1}{i \partial^+}\left[\left( i \sig^i \partial^i+e \sig^i A^i
 +i m_e\right)\xi_e\right]
 \nn\\
 &&~~~~~~+\xi_p^\dagger\left(i\hspace{.1mm}\sigma^i\partial^i-
 e\hspace{.1mm}\sig^iA^i-i\hspace{.1mm} m_p\right)
 \frac{1}{i \partial^+}\left[\left( i \sig^i \partial^i-e \sig^i A^i
 +i m_p\right)\xi_p\right]\nn\\
&&~~~~~~ -\frac{1}{2}J^+\frac{1}{\left(\partial^+\right)^2} J^+
 +J^+ \frac{\partial^i}{\partial^+} A^i
\;.\label{eq:can2}
\eea
Note that $i=1,2$ only; 
$J^+=2 e \left(
\xi_p^\dagger \xi_p-\xi_e^\dagger \xi_e
\right)$, and $\sig^i$ are  the standard $SU(2)$ Pauli matrices.
The dynamical fields are $A^i$, $\xi_e$ and $\xi_p$, the transverse photon
and two-component electron and proton fields respectively. 
 For the relation between $\psi$ and $\xi$ and a comprehensive
  summary of our light-front conventions
see Appendix~A. 

The free Hamiltonian 
is  
\bea
&&h=\left.H\right|_{(e=0)}=
 \int_p \sum_s \left(b_s^\dagger(p) b_s(p)~\frac{{p^\perp}^2+m_e^2}{p^+}+
 B_s^\dagger(p) B_s(p)~
 \frac{{p^\perp}^2+m_p^2}{p^+}+
  a_s^\dagger(p) a_s(p)~
 \frac{{p^\perp}^2}{p^+}\right)
\;,\label{eq:freebaby}
\eea
plus the anti-fermions.
The notation for our free spectrum is $h |i\rangle = \veps_i |i\rangle$
with $\sum_i |i\rangle\langle i| = 1$, where the sum over $i$ implies a sum over all Fock
sectors, momenta, and spin.
Next, we normal-order all interactions and neglect zero modes.
The canonical
 interactions from Eq.~(\ref{eq:can2}) that  we use in this paper are
\bea
v_{1e}&=& \int d^2 x^{\perp}dx^- ~{\cal V}_{1e}~\;,\;~
v_{1p}= \int d^2 x^{\perp}dx^- ~{\cal V}_{1p}~\;,\;~
v_{2}= \int d^2 x^{\perp}dx^- ~{\cal V}_2
\;,
\label{eq:rtrt}
\eea
with
\bea
{\cal V}_{1e}&=&
e\xi_e^\dagger
 \hspace{.1mm}\sig^iA^i
 \frac{1}{i \partial^+}\left[\left( i \sig^i \partial^i
 +i m_e\right)\xi_e\right]+
 e\xi_e^\dagger\left(i\hspace{.1mm}\sigma^i\partial^i
 -i\hspace{.1mm} m_e\right)
 \frac{1}{i \partial^+}\left[  \sig^i A^i
 \xi_e\right]-
 2 e \xi_e^\dagger \xi_e
 \frac{\partial^i}{\partial^+} A^i
\label{eq:eemission}\;,\\
{\cal V}_{1p}&=&
 -e\xi_p^\dagger
 \hspace{.1mm}\sig^iA^i
 \frac{1}{i \partial^+}\left[\left( i \sig^i \partial^i
 +i m_p\right)\xi_p\right]-e
 \xi_p^\dagger\left(i\hspace{.1mm}\sigma^i\partial^i-i\hspace{.1mm} m_p\right)
 \frac{1}{i \partial^+}\left[ \sig^i A^i
 \xi_p\right]
 +
 2 e
\xi_p^\dagger \xi_p \frac{\partial^i}{\partial^+} A^i
\label{eq:pemission}\;,\\
{\cal V}_2&=&
-\frac{1}{2}J^+\frac{1}{\left(\partial^+\right)^2} J^+
\label{eq:instant}\;.
\eea
These are the photon emission and absorption by the electron, photon emission and absorption
by the proton, and
instantaneous photon interactions respectively. 

Given the canonical Hamiltonian, $H$, we cut off the theory by 
requiring the free energies of all states to satisfy
\bea
&& \veps_i \leq
 \frac{{{\cal P}^\perp}^2+\Lam^2}{{\cal P}^+}
 \;,
 \eea
 where $\Lam$ is the bare cutoff, and ${\cal P}=
 \left({\cal P}^+,{\cal P}^\perp\right)$ is the total momentum of
 the hydrogen state.
 Then, with a Bloch transformation
  we remove the states with free energies satisfying
 \bea
&&\frac{{{\cal P}^\perp}^2+\lam^2}{{\cal P}^+}\leq  \veps_i \leq
 \frac{{{\cal P}^\perp}^2+\Lam^2}{{\cal P}^+}
 \;,
 \eea
 where $\lam$ is the effective cutoff.
 The result is an effective Hamiltonian, $H_\lam$, acting in 
 the low energy ($ \veps_i \leq
 \frac{{{\cal P}^\perp}^2+\lam^2}{{\cal P}^+}$) space alone.
 We do not discuss the derivation of $H_\lam$ any further, but 
 instead refer the interested reader to
 Appendix~B.
 
 Given $H_\lam$, we then  make the following division
 \bea
 &&H_\lam={\cal H}_o + \left(H_\lam-{\cal H}_o\right)~\equiv~{\cal H}_o+{\cal V}~,
 \eea
 where ${\cal H}_o$ is an approximation that can be solved non-perturbatively
 (for this QED calculation) and ${\cal V}$ is treated in BSPT. The 
 test of ${\cal H}_o$ is whether BSPT converges or not and closely related: is 
 the $\lam$-dependence of the spectrum weakened by higher orders of BSPT?
\vskip.2in
\noindent
\centerline{{\bf III. LOWEST ORDER SCHR\"{O}DINGER EQUATION}}
\vskip.2in

The 
primary {\em assumption} we make in this QED bound-state calculation is
that the Coulomb interaction dominates all other physics. In this work we
will treat the Coulomb interaction between the electron and proton to 
all orders in all Fock sectors.
After this assumption,
 the kinematic length scale of our system is fixed,
\beaa
&&a_o \sim \frac{1}{p} \sim \frac{1}{m \alpha}\sim \frac{137}{m}
\;,
\eeaa
which then fixes our dynamical time and length scale,
\beaa
&&t\sim\frac{1}{p^2/(2m)}\sim\frac{1}{m \alpha^2}\sim \frac{137^2}{m}
\;:
\eeaa
as is well known,
dynamical changes occur very slowly in this system.
Note that in this QED calculation we will treat photons as free since they 
carry no charge and interact very weakly at low energies. This of course
changes drastically for QCD since gluons do carry color charge and interact
strongly at low energies.
After choosing ${\cal H}_o$,
 the $\alpha$-scaling of our BSPT is fixed, and the spectrum is then
calculated to some desired order in $\alpha$ and $\log\alpha$.

In the Coulomb gauge, the Coulomb interaction appears directly in the canonical Hamiltonian,
which 
of course is not true in the light-cone gauge.\footnote{However,
 a confining potential does appear directly in the canonical Hamiltonian in
the light-cone gauge, which is a convenient starting point for  $QCD_{3+1}$ \cite{perrybrazil}.}
In the light-cone gauge,
the Coulomb interaction arises from a combination of two types of interactions
in our effective Hamiltonian, instantaneous photon exchange and the two time
orderings of dynamical photon exchange. Graphically this is shown in Fig.~1.
These interactions arise from first and second-order effective interactions
respectively. See Eq.~(\ref{eq:heff}) of Appendix~B for 
the form of the effective Hamiltonian, $H_\lam$.

The time-independent
Schr{\"o}dinger equation in light-front coordinates 
that the sum of the three time-ordered diagrams in
Fig.~1  satisfies is \footnote{For a derivation of Eq.~(\ref{eq:lowsch}) 
 from the Schr{\"o}dinger equation in Fock space 
see Eqs.~(81) to (83) in \S III.B.1 of Ref.~\cite{billy}.}
\bea
 \left({\cal M}_{_{N}}^2-\frac{{\kap^\pr}^2+m_e^2}{x^\pr}- \frac{{\kap^\pr}^2+m_p^2}{1-x^\pr}
 \right) {\tilde{\phi}}_N(x^\pr \kap^\pr s_e^\pr s_p^\pr)
 &=&
 \sum_{s_e s_p}\int d^2 \kap/(2 \pi)^2 \int_0^1 dx/(4 \pi)\nn\\
 &&~~~~~~~~~~\times~
 {\tilde V}_c~ {\tilde{\phi}}_N(x \kap s_e s_p)
 \;.\label{eq:lowsch}
 \eea
 ${\cal M}_{_{N}}^2$ is the  mass squared eigenvalue of the state ${\tilde \phi}_N$,
where ``$N$" labels
 all the quantum numbers of this state.
 The tildes will be notationally convenient below. 
We have introduced the following Jacobi variables
\bea
p_e&=&(x {\cal P}^+ , \kappa+ x {\cal P}^\perp)~,\\
p_e^\pr&=&(x^\pr {\cal P}^+ , \kappa^\pr+ x^\pr {\cal P}^\perp)
\;,
\eea
where $p_e$ and $p_e^\pr$ are the initial and final electron three-momentum
respectively, and
\bea
&&p_e+p_p=p_e^\pr+p_p^\pr={\cal P}=({\cal P}^+,{\cal P}^\perp)
\;
\eea
is the total momentum of the hydrogen state.
Note that $\kap$ is a two-vector.
 The norm is defined by
 \bea
 \sum_{s_e s_p}\int d^2 \kap/(2 \pi)^2 \int_0^1 dx/(4 \pi)~
 {{\tilde{\phi}}_N}^\ast(x \kap s_e s_p) 
 {\tilde{\phi}}_{N^\pr}(x \kap s_e s_p)\;=\;\del_{N N^\pr}\;.
 \eea
${\tilde V}_c$  is the sum of the interactions given by the
three diagrams in Fig.~1, and 
will not be written in all its gory detail.\footnote{The interested reader should consult
Eqs.~(70) and (71) and the discussion below in \S III.A.2 of Ref.~\cite{billy};
these equations are for the equal mass case, but are readily generalized
to the unequal mass case; also note that in this reference we used a similarity
transformation instead of a Bloch transformation; the Bloch transformation was chosen
for the current paper because of its simplicity.} The leading order term of
${\tilde V}_c$ in a non-relativistic expansion is defined as $V_c$ and is written below.

The non-relativistic expansion is defined in the following way.
A coordinate change which takes the range of longitudinal momentum fraction, $x \in [0,1]$ to 
 $\kap_z \in [-\infty, \infty]$ is defined:
 \bea
 x&=&\frac{\kap_z+\sqrt{\kap^2+\kap_z^2+m_e^2}}{\sqrt{\kap^2+\kap_z^2+m_e^2}+
 \sqrt{\kap^2+\kap_z^2+m_p^2}}\;.
 \label{eq:coord}\eea
 This step can be taken for relativistic kinematics, but
 there may be no advantage. 
Then, the non-relativistic expansion is an expansion in $\bp/m$; i.e.,
we assume
\bea
 &&m\gg|{\bf p}|\;,\label{eq:nr}
\eea
where we have defined a
new three-vector in terms of our transverse Jacobi variable, $\kap$, and our
new longitudinal momentum variable, $\kap_z$, which replaces our longitudinal momentum
fraction, $x$,
  \bea
 {\bf p}&\equiv& (\kap,\kap_z)
 \;.
 \label{eq:coord2}\eea
 Note that the free mass squared in the Schr{\"o}dinger equation,
Eq.~(\ref{eq:lowsch}), after this coordinate change, becomes
\bea
 \frac{{\kap}^2+m_e^2}{x}+ \frac{{\kap}^2+m_p^2}{1-x}
 &=&\left(\sqrt{m_e^2+{\bf p}^2}+\sqrt{m_p^2+{\bf p}^2}~\right)^2\nn\\
 &=&
 (m_e+m_p)^2+2 (m_e+m_p) \left(
 \frac{{\bf p}^2}{2 m}-\frac{{\bf p}^4 (m_e-m_p)^2}{8\hspace{.5mm} m\hspace{.5mm}
  m_e^2 m_p^2}+{\cal O}\left(\frac{\bp^6}{m^5}\right)
 \right)
 \;,\label{eq:30}
 \eea
 which is invariant under rotations in the space of vectors ${\bf p}$---but not invariant
 under $p_z$ boosts. Here we begin to see 
 longitudinal boost invariance being replaced by an expanded kinematic
  rotational invariance in the theory.
 $m$ is the reduced mass given in Eq.~(\ref{eq:reduced}).

Now note that the leading order term of ${\tilde V}_c$ in an expansion in $\bp/m$ is contained in
\bea
{\tilde V}_c&\sim& (m_e+m_p)^2 
\left(-\frac{4 e^2}{q_z^2}+
\frac{4 e^2 {q^\perp}^2}{q_z^2 {\bf q}^2}\theta_{H}
\right)\del_{s_e s_e^\pr} \del_{ s_ps_p^\pr}
\theta_{L}
\;,
\eea
where
\bea
{\bf q} &=& {\bf p}^\pr-{\bf p}\\
\theta_{L}
&=&
\theta\left(\lam^2-
\left(\sqrt{m_e^2+{\bf p}^2}+\sqrt{m_p^2+{\bf p}^2}~\right)^2\right)
\theta\left(\lam^2-
\left(\sqrt{m_e^2+{{\bf p}^\pr}^2}+\sqrt{m_p^2+{{\bf p}^\pr}^2}~\right)^2\right)\;,\\
\theta_{H}&=&\theta
\left(\left((m_e+m_p)^2+2 (m_e+m_p) \frac{{\bf q}^2}{2|q_z|}\right)-\lam^2
\right)\nn\\
&&~~~~~\times~
\theta
\left(\Lam^2-\left((m_e+m_p)^2+2 (m_e+m_p) \frac{{\bf q}^2}{2|q_z|}\right)
\right)
\;.
\eea
Note that $\theta_L$ and $\theta_H$ 
 are the constraints that arise from the Bloch transformation.

It is convenient to define new cutoffs which subtract off the total free
constituent masses of the state
\bea
\tilde{\lam}&\equiv&\lam-(m_e+m_p)~,\label{eq:21}\\
\tilde{\Lam}&\equiv&\Lam-(m_e+m_p)
\;.
\eea
In the limit $m_p \rightarrow \infty$ we require $\tilde{\lam}$ and $\tilde{\Lam}$
to be held fixed.
Note that this implies
\bea
\frac{\lam^2-(m_e+m_p)^2}{2(m_e+m_p)}&=&\tilde{\lam}+\frac{\tilde{\lam}^2}{2(m_e+m_p)}
\stackrel{(m_p \rightarrow \infty)}{\longrightarrow}\tilde{\lam}~,\\
\frac{\Lam^2-(m_e+m_p)^2}{2(m_e+m_p)}&=&\tilde{\Lam}+\frac{\tilde{\Lam}^2}{2(m_e+m_p)}
\stackrel{(m_p \rightarrow \infty)}{\longrightarrow}\tilde{\Lam}
\;.\label{eq:21p}
\eea

In terms of these new cutoffs,
 $\theta_{L}$ and $\theta_{H}$ above become
\bea
\theta_{L}&=& \theta
\left(\tilde{\lam}-\frac{{\bf p}^2}{2 m}+{\cal O}\left(\frac{\bp^4}{m^3}\right)\right)
\theta
\left(\tilde{\lam}-\frac{{{\bf p}^\pr}^2}
{2 m}+{\cal O}\left(\frac{|{\bf p}^\pr|^4}{m^3}\right)\right)~,\\
\theta_{H}&=&\theta\left(\frac{{\bf q}^2}{2|q_z|}-\tilde{\lam}\right)
\theta\left(\tilde{\Lam}-\frac{{\bf q}^2}{2|q_z|}\right)
\;.
\eea

To see the Coulomb interaction arising from the $|e p\rangle$ sector alone,
we make the following requirements (which are motivated from the previous two
equations)
\bea
\frac{\bp^2}{m}\ll \tilde{\lam} \ll \bp~~~~~~~~~~~&{\rm and}&~~~~~~~~~~~
\tilde{\Lam}\gg \bp
\label{eq:constraints}
\;,
\eea
also demanded for $|{\bf p}^\pr|$ of course.
 These constraints
 will be maintained consistently in this paper. 
Given these restrictions we have
\bea
\theta_L &\approx& 1~,\\
\theta_H &\approx& 1
\;.
\eea
 ${\tilde V}_c$ becomes
\bea
{\tilde V}_c&\sim& V_c
\;,
\eea
where
\bea
V_c
&\equiv&
(m_e+m_p)^2 
\left(-\frac{4 e^2}{q_z^2}+
\frac{4 e^2 {q^\perp}^2}{q_z^2 {\bf q}^2}
\right)\del_{s_e s_e^\pr} \del_{s_p s_p^\pr}\nonumber\\
&=&-(m_e+m_p)^2 \left(\frac{4 e^2}{{\bf q}^2}\right)\del_{ s_es_e^\pr} \del_{s_p s_p^\pr}~~.
\eea

To finish showing how the Coulomb interaction arises in a light-front 
Hamiltonian approach,
 we need to know the Jacobian of the
coordinate transformation of Eq.~(\ref{eq:coord}), 
\bea
 J(p)&=&\frac{dx}{d\kap_z}=
 \frac{\left(\kap_z+\sqrt{{\bf p}^2+m_e^2}\right)\left(\sqrt{{\bf p}^2+m_p^2}-\kap_z\right)}
 {\sqrt{{\bf p}^2+m_e^2}\sqrt{{\bf p}^2+m_p^2}\left(
 \sqrt{{\bf p}^2+m_e^2}+\sqrt{{\bf p}^2+m_p^2}~\right)}\nn
 \\
 &=&
  \frac{1}{m_e+m_p}\left(1+\kap_z\left(\frac{1}{m_e}-\frac{1}{m_p}\right)-
\frac{\left({\bf p}^2+2\kap_z^2\right)}{2 m_e m_p}+
{\cal O}\left(\frac{\bp^3}{m^3}\right)\right)
 \;.
 \eea
It is also convenient to redefine the norm
\bea
\del_{N N^\pr}&=&\sum_{s_e s_p}\int d^2 \kap/(2 \pi)^2 \int_0^1 dx/(4 \pi)~
 {\tp_N}^\ast(x \kap s_e s_p) \tp_{N^\pr}(x \kap s_e s_p) \nn\\
 &=&
 \sum_{s_e s_p} \int d^3 p~ J(p)/(16\pi^3)~
 {\tp_N}^\ast({\bf p} s_e s_p) \tp_{N^\pr}({\bf p} s_e s_p)
\nn \\
 &\equiv&\sum_{s_e s_p} \int d^3 p \;\phi_N^{ \ast}({\bf p} s_e s_p) \phi_{N^\pr}({\bf p} s_e s_p)
 \;.\label{eq:norm}\eea
In this last line the tildes are removed from the wave functions by defining
\bea
\phi_{N}({\bf p} s_e s_p)&\equiv&\sqrt{\frac{J(p)}{16 \pi^3}}\tp_{N}({\bf p} s_e s_p)~.
\label{eq:jac}
\eea

Putting it all together,  the leading order expression for Eq.~(\ref{eq:lowsch})
in an expansion in $\bp/m$ given the restrictions of Eq.~(\ref{eq:constraints}) is
\bea
 \left(-\beta_{n}+\frac{{{\bf p}^\pr}^2}{2m}
 \right) \phi_N({\bf p}^\pr  s_e^\pr s_p^\pr)&=&
 \frac{\alpha}{2 \pi^2}\int\frac{ d^3 p }{({\bf p}-{\bf p}^\pr)^2}
  \phi_N({\bf p} s_e^\pr s_p^\pr)
 \;,\label{eq:sch}
\eea
which we see is the  non-relativistic Schr{\"o}dinger equation of hydrogen.
$m$ is the reduced mass and $-\beta_n$ is the binding energy defined by
\bea
\beta_n&=&\frac{{\cal M}_{n}^2-(m_e+m_p)^2}{2 (m_e+m_p)}
\;.\label{eq:beta}
\eea
The well known bound spectrum is
\bea
\beta_n&=&-\frac{Ryd}{n^2}
\;,
\eea
where $Ryd = m \alpha^2/2$ of course.
Note that Eq.~(\ref{eq:sch}) fixes the $\alpha$-scaling of $\bp$:
\bea
&&\bp\sim m\alpha~.
\eea
Thus we see that the restrictions of Eq.~(\ref{eq:constraints}) become
\bea
m \alpha^2 \ll \tilde{\lam} \ll m \alpha~~~~~~~~~~~&{\rm and}&~~~~~~~~~~~
\tilde{\Lam}\gg m \alpha
~,
\eea
which is consistent with Eq.~(\ref{eq:2cv}) as advertised earlier.
\vskip.2in
\noindent
\centerline{{\bf IV. LAMB SHIFT CALCULATION}}
\vskip.2in

Given our lowest order spectrum, we  
proceed with BSPT. As advertised, this will be divided into {\em low} and {\em high} 
intermediate photon
energy  calculations. Before proceeding with these respective
calculations, we  discuss whether   Coulomb exchange  can be treated perturbatively or 
non-perturbatively   in the
respective regions.

For the low energy intermediate photon, the Coulomb interaction between the intermediate
electron and proton must be treated non-perturbatively, whereas this interaction
can be treated perturbatively for the high energy intermediate photon contribution.
 This is seen
by noting that each additional Coulomb exchange contributes a Coulomb matrix element and an
energy denominator which is dominated by the larger photon energy scale. Thus each additional
Coulomb exchange contributes
\bea
&&\frac{\langle \frac{\alpha}{|{\bf r}|} \rangle}{|{\bf k}|}\leq
\frac{m \alpha^2}{|{\bf k}|_{min}}
\;.\label{eq:ex}
\eea
For the low energy photon contribution, in principle
$|{\bf k}|_{min}=0$,
 and each additional Coulomb exchange can
contribute ${\cal O}(1)$, and therefore must be treated non-perturbatively. 
Of course, when the Coulomb interaction is treated non-perturbatively, low-energy
intermediate protons and electrons form bound states from which
long-wavelength photons decouple. This non-perturbative effect leads to 
$|{\bf k}|_{min}\sim 16.64~{\rm Ryd}$; see Eq.~(\ref{eq:ttbbg}) below.
For 
the high energy photon contribution, $|{\bf k}|_{min}=\tilde{\lam}$ and from Eq.~(\ref{eq:32415}) each
additional Coulomb exchange thus contributes at  most
\bea
&&\frac{m \alpha^2}{\tilde{\lam}}= \sqrt{\alpha} \sim 8.5~\times~10^{-2}\label{eq:exhigh}
\;,
\eea
and can therefore be treated perturbatively.

\vskip.2in
\noindent
\centerline{{\bf A. Low energy contribution}}
\vskip.2in

The low energy shift arises from two sources  which are shown in Fig.~2. The first 
term comes from the low energy photon emission part of the effective Hamiltonian,
$
\langle a | v_{1e} | b \rangle
$,
treated in second-order BSPT. Recall Eqs.~(\ref{eq:rtrt}) and 
(\ref{eq:eemission}) for the form of $v_{1e}$.\footnote{
Note that the term where the proton emits and subsequently absorbs a photon is down by two powers of the
proton mass with respect to the term where the electron emits and absorbs a photon. This result
is subtle though, because it is true only after the
 light-front infrared divergences have canceled
between   two  diagrams analogous to the ones in Fig.~2.}
The second term is the result of renormalizing the one loop electron self-energy: a counterterm
is added to the second-order self-energy effective interaction in 
$
\langle a |H_\lam | b \rangle
$, which results
in a finite (except for infrared divergences) shift to the electron self-energy.
This is shown in Fig.~3. 
 The counterterm is  fixed by requiring the electron self-energy
  to evolve coherently with
the cutoff.
The details of defining this counterterm, for the equal mass case,
 can be found in \S III.A.1 of
Ref.~\cite{billy}. A discussion of the physical ideas behind coupling coherence can
be found in Ref.~\cite{perry2}.
 For
  further comments on coupling coherence, see the paragraph containing Eq. (39) of 
  Ref. \cite{billy}.  
 
Before proceeding with the calculation, we define the binding energy of hydrogen, $-B_N$, in
terms of the mass-squared, $M_N^2$,
\bea
M_N^2&=&(m_e+m_p+B_N)^2
\;.
\eea
Assuming $B_N$ is finite as $m_p \rightarrow \infty$ we have
\bea
\frac{M_N^2-(m_e+m_p)^2}{2(m_e+m_p)}&=&B_N+\frac{B_N^2}{2(m_e+m_p)}
\stackrel{(m_p \rightarrow \infty)}{\longrightarrow}B_N
\;.
\label{eq:4444}
\eea
Recalling Eq.~(\ref{eq:beta}), which is 
the definition of the zeroth order binding, $-\beta_n$, in terms of the
zeroth order mass-squared, ${\cal M}_n^2$; and also defining the mass
squared corrections, $\del M_N^2$,  by
\bea
 M_N^2&=& {\cal M}_n^2+\del M_N^2 
\;,
\eea
combined with Eq.~(\ref{eq:4444}),
gives 
\bea
B_N&=&\beta_n+\frac{\del M_N^2}{2 (m_e+m_p)} \stackrel{(m_p \rightarrow \infty)}{\longrightarrow}
\beta_n+\frac{\del M_N^2}{2 m_p}\label{eq:robert}
\;.
\eea
Defining the binding corrections, $-\del B_N$, by
\bea
  B_N &=& \beta_n+\del B_N
\;,
\eea
combined with Eq.~(\ref{eq:robert}), gives
\bea
\del B_N&=&\frac{\del M_N^2}{2 m_p}
\;,
\label{eq:binding}
\eea
a useful formula to be used below. This formula is useful because $\del M_N^2$ is 
calculated below, but 
$\del B_N$ is the quantity that is measured.

The low energy calculation proceeds as follows. 
The first term of Fig.~2 is a second-order BSPT
shift which contributes the following to the mass squared eigenvalue:
\bea
\del M_{L1}^2&=&\sum_{N^\pr} \int_k \sum_{s_\gamma}\frac{
\left|\langle \psi_N\left({\cal P}\right) |
v_{1e}a_{s_\gamma}^\dagger(k)|\psi_{N^\pr}\left(
{\cal P}-k\right)\rangle
\right|^2 \theta_{L1}
}
{DEN_1 (Vol)^2}
\;,\label{eq:L1}
\eea
where $k$ and $s_\gamma$ are the photon's three-momentum and spin respectively, 
${\cal P}=\left({\cal P}^+,{\cal P}^\perp\right)$ is the total momentum of
the hydrogen state $\psi_N$, and 
$v_{1e}$ is the photon emission interaction given in Eq.~(\ref{eq:rtrt}).
$\theta_{L1}$  restricts  the
 energies of the initial, intermediate and final states to be below the effective
 cutoff $\frac{\lam^2+{{\cal P}^\perp}^2}{{\cal P}^+}$.
 The explicit form of these restrictions is discussed below.
Continuing the description of Eq.~(\ref{eq:L1}), 
\bea
\int_k &&= \int \frac{d^2k^\perp dk^+ \theta(k^+)}
 {16 \pi^3 k^+}=\int \frac{d^3 k}{ (2 \pi)^3 (2 |{\bf k}| )}
\;.
\eea
The last step comes from recalling that for a photon $k^+=k^0+k^3=|{\bf k}|+k^3$.
 The denominator and volume factors are
\bea
Vol&=&\int d^2 x^{\perp}dx^-=16 \pi^3 ~\del^3\left({\cal P}-{\cal P}\right)\;,\\
DEN_1&=&{\cal P}^+ \left(
\frac{{{\cal P}^\perp}^2+{\cal M}_n^2}{{\cal P}^+}
-\frac{{({\cal P}-k)^\perp}^2+{\cal M}_{n^\pr}^2}{({\cal P}-k)^+}
-\frac{{k^\perp}^2}{k^+}
\right)\label{eq:den1}
\;.
\eea
The two-body states  are
\bea
|\psi_N\left({\cal P}\right)\rangle&=&\int_{p_e p_p} \sqrt{p_e^+ p_p^+} 16 \pi^3 \del^3 
\left( {\cal P} -p_e-p_p\right)\tilde{\phi}_N(p_e p_p) b_{s_e}^\dagger(p_e) B_{s_p}^\dagger
(p_p) | 0 \rangle~,\\
|\psi_{N^\pr}\left({\cal P}-k\right)\rangle&=&\int_{k_1 k_2} \sqrt{k_1^+ k_2^+} 16 \pi^3 \del^3 
\left( {\cal P}-k -k_1-k_2\right)\tilde{\phi}_{N^\pr}(k_1 k_2) b_{s_e^\pr}^\dagger(k_1) 
B_{s_p^\pr}^\dagger
(k_2) | 0 \rangle
\;,
\eea
where  $\phi_N$ are solutions to Eq.~(\ref{eq:sch}), 
 the non-relativistic Schr{\"o}dinger equation of hydrogen,
 and ${\tilde{\phi}}_N$ is related to $\phi_N$ by Eq.~(\ref{eq:jac}).

Straightforward algebra leads to
\bea
\del M_{L1}^2&=&{\sum_{N^\pr}}^c \int_k \int_{p_e} \theta\left({\cal P}^+ - p_e^+\right)
\int_{p_e^\pr}\theta\left({\cal P}^+ - {p_e^\pr}^+\right) \int_{k_1 k_3} 
 \left(p_e^+{ p_e^\pr}^+ k_1^+ k_3^+\right) 
 \nn\\
 &&~~~~~\times\left(16 \pi^3 \del^3 (k+k_3-p_e)\right)\left(16 \pi^3 \del^3 (k+k_1-p_e^\pr)\right) 
 \tilde{\phi}_N^\ast\left(p_e^\pr,{\cal P}-p_e^\pr\right)
\nonumber\\
&&~~~~~\times\tilde{\phi}_{N^\pr}\left(k_1,{\cal P}-k-k_1\right)
\tilde{\phi}_{N^\pr}^\ast\left(k_3,{\cal P}-k-k_3\right)
\tilde{\phi}_N\left(p_e,{\cal P}-p_e\right) \frac{N_1\theta_{L1}}{DEN_1}
\;,\label{eq:l1}
\eea
where $N$ and $N^\pr$ are shorthands for $(n,l,m_l)$ and $(n^\pr,l^\pr,m_l^\pr)$ respectively,
the usual principal and angular momentum quantum 
  numbers of non-relativistic hydrogen.
The ``$c$" on the sum emphasizes the fact that the continuum states must be included also.
See Eq.~(\ref{eq:den1}) for $DEN_1$.
 $N_1$ is given by 
 \bea
 N_1&=&
 \sum_{s_e^\pr s_\gam}
 \frac{\langle 0 | b_{s_e}(p_e^\pr) ~v_{1e}~ b_{s_e^\pr}^\da(k_1) a_{s_\gam}^\da(k)
 |0\rangle \langle 0| b_{s_e^\pr}(k_3) a_{s_\gam}(k) ~v_{1e}~ b_{s_e}^\da(p_e) |0\rangle}
 {\sqrt{p_e^+ {p_e^\pr}^+ k_1^+ k_3^+}
 \left(16 \pi^3 \del^3 (k+k_3-p_e)\right)\left(16 \pi^3 \del^3 (k+k_1-p_e^\pr)\right)}
 \;
 \eea
(for $v_{1e}$ see Eq.~(\ref{eq:rtrt})), which after some algebra becomes
\bea
N_1&=& (4 \pi \alpha) \left[
2 m_e^2\left(\frac{1}{p_e^+}-\frac{1}{k_3^+}\right)
\left(\frac{1}{{p_e^\pr}^+}-\frac{1}{k_1^+}\right)
\right.\nn\\
&&~~~~~~~~~~+\left.
\left(\frac{2 k^i}{k^+}-\frac{k_1^i(s_e)}{k_1^+}-\frac{{p_e^\pr}^i(\overline{s}_e)}{{p_e^\pr}^+}\right)
\left(\frac{2 k^i}{k^+}-\frac{p_e^i(s_e)}{p_e^+}-\frac{k_3^i(\overline{s}_e)}{k_3^+}\right)
\right]\label{eq:N1}
\;,
\eea
where we have defined a new object,
\bea
p^i(s)&=&p^i+i~s~ \eps_{ij} ~ p^j
\;.
\eea
Notation:  $i= 1,2$ only, $s= \pm 1$ only, $\overline{s} = -s$, $\eps_{12}=-\eps_{21}=1$ and 
$\eps_{11}=\eps_{22}=0$.

We now discuss $\theta_{L1}$ and then simplify $\del M_{L1}^2$ further.
Recall Eqs.~(\ref{eq:30}) and (\ref{eq:21}). We 
see that 
after the coordinate change defined by Eq.~(\ref{eq:coord}), in the $m_p \rightarrow \infty$
limit, 
\bea
\theta_{L1}&=&\theta\left(
\tilde{\lam}-T_1
\right)\theta\left(\tilde{\lam}-\frac{{\bf p}^2}{2m}+{\cal O}(\alpha^4)\right)
\theta\left(\tilde{\lam}-\frac{{{\bf p}^\pr}^2}{2m}+{\cal O}(\alpha^4)\right)
~,\\
T_1&=&|{\bf k}| + \sqrt{({\bf p}-{\bf k})^2+m_e^2}-m_e 
\;,\label{eq:t1}
\eea
where we have used the fact that
the wave functions  restrict $|{\bf p}| \sim m \alpha$. Recall that  we are always assuming
$m \alpha^2 \ll \tilde{\lam}\ll m \alpha$. Thus, $\theta_{L1}$ can be simplified: 
\bea
\theta_{L1}&\approx&\theta\left(
\tilde{\lam}-T_1
\right)
\;.
\eea
From the form of Eq.~(\ref{eq:t1}), we see that this constrains
 the photon momentum to satisfy 
\bea
|{\bf k}|&\leq& \tilde{\lam}\;,\label{eq:lowcon}
\eea
to leading order in $\alpha$.

Note that the constraints coming from $\theta_{L1}$, summarized by
Eq.~(\ref{eq:lowcon}), require the photon momenta in $\del M_{L1}^2$ 
of Eq.~(\ref{eq:l1})
to satisfy
\bea
k&\ll& p_e,p_e^\pr
\;.
\eea
Thus, Eq.~(\ref{eq:l1}) can be simplified further,
\bea
\del M_{L1}^2&\approx&{\sum_{N^\pr}}^c \int_k \int_{p_e} \theta\left({\cal P}^+ - p_e^+\right)
\int_{p_e^\pr}\theta\left({\cal P}^+ - {p_e^\pr}^+\right) 
 \left(p_e^+{ p_e^\pr}^+ \right) \tilde{\phi}_N^\ast\left(p_e^\pr,{\cal P}-p_e^\pr\right)
 \nn\\
 &&~~~~~\times 
  \tilde{\phi}_{N^\pr}\left(p_e^\pr,{\cal P}-p_e^\pr\right)
\tilde{\phi}_{N^\pr}^\ast\left(p_e,{\cal P}-p_e\right)
\tilde{\phi}_N\left(p_e,{\cal P}-p_e\right) \left.\frac{N_1}{DEN_1}\right|_{(k_3=p_e,k_1=p_e^\pr,
|{\bf k}| \leq \tilde{\lam})}
\;.\label{eq:l1sec}
\eea

In the  $m_p \longrightarrow \infty$
limit, ${\cal P}^+\longrightarrow m_p$, and $DEN_1$ becomes
\bea
DEN_1&=&2 m_p\left(\beta_n-\beta_{n^\pr}-|{\bf k}|\right)\left(
1+{\cal O}\left(1/m_p\right)\right)
\;,
\eea
where we have used $\frac{{k^\perp}^2}{k^+}+k^+=2|{\bf k}|$, 
valid for an on-mass-shell photon (all particles in a Hamiltonian approach
are on-mass-shell).
$-\beta_n$ is the binding energy of non-relativistic hydrogen defined in Eq.~(\ref{eq:beta}),
with numerical value $Ryd/n^2$ for the bound-states.

In the region of integration, $|{\bf k}| \leq \tilde{\lam}= m \alpha \sqrt{\alpha}\ll |{\bf p}|$, so
after the coordinate change
of Eq.~(\ref{eq:coord})  (recall Eq.~(\ref{eq:coord2})) we have
\bea
\frac{N_1}{4 \pi \alpha}&=&\frac{4 {k^\perp}^2}{{k^+}^2}+\frac{4 {k^\perp}^2}{k^+ m}+
\frac{4 ~{\bf p}^\perp\cdot {\bf p}^{\pr \perp}}{m^2} \nn\\
&&~~~~~-\frac{4k^\perp}{k^+ m}\cdot\left(  {\bf p}^\perp+{\bf p}^{\pr \perp}
-\frac{{\bf p}^\perp p_z}{m}-\frac{{\bf p}^{\pr \perp}p^\pr_z}{m} \right)+{\cal O}
\left(\alpha^2 \sqrt{\alpha}\right)
\;.
\eea
The rest of the integrand is even under $k^\perp \rightarrow - k^\perp$, so these terms in the last
line, odd in $k^\perp$, do not contribute. 

Putting it all together,
recalling Eq.~(\ref{eq:binding}),
 we have
\bea
\del B_{L1}&=&\frac{\del M_{L1}^2}{2 m_p}\approx\frac{\alpha}{4 \pi^2 }
 {\sum_{N^\pr}}^c \int \frac{d^3 k}{  |{\bf k}| }
~\theta\left(\tilde{\lam}-|{\bf k}|\right)
\int d^3 p \int d^3 p^\pr
\phi_N^\ast\left({\bf p}^\pr\right)
\phi_{N^\pr}\left({\bf p}^\pr\right)\nn\\
&&~~~~~~~~~~~~~~~~~~~~~~~~~~~~~~\times
\phi_{N^\pr}^\ast\left({\bf p}\right)\phi_N\left({\bf p}\right)
\frac{\frac{ {k^\perp}^2}{{k^+}^2}+\frac{ {k^\perp}^2}{k^+ m}+
\frac{{\bf p}^\perp\cdot {\bf p}^{\pr \perp}}{m^2}}
{\beta_n-\beta_{n^\pr}-|{\bf k}|}
\;,\label{eq:bl1}
\eea
where we recalled Eq.~(\ref{eq:jac}), the relation between
$\phi_N$ and $\tilde{\phi}_N$.
This is infrared ($k^+ \rightarrow 0$) divergent, but we must add diagram~L2 of Fig.~2
 to get
the total low-energy shift. 

As previously mentioned,
Diagram~L2 of Fig.~2 arises from the sum of an effective second-order 
electron self-energy interaction and 
a counterterm defined such that the electron self-energy runs coherently. 
 The result of this interaction is to add the following
to the binding
\bea
\del B_{L2}&=&\frac{\del M_{L2}^2}{2 m_p}=-\frac{\alpha}{4 \pi^2 }
 {\sum_{N^\pr}}^c \int \frac{d^3 k}{  |{\bf k}| }
~\theta\left(\tilde{\lam}-|{\bf k}|\right)
\int d^3 p \int d^3 p^\pr
\phi_N^\ast\left({\bf p}^\pr\right)
\phi_{N^\pr}\left({\bf p}^\pr\right)\nn\\
&&~~~~~~~~~~~~~~~~~~~~~~~~~~~~~~\times
\phi_{N^\pr}^\ast\left({\bf p}\right)\phi_N\left({\bf p}\right)
\frac{\frac{ {k^\perp}^2}{{k^+}^2}+\frac{ {k^\perp}^2}{k^+ m}+
\frac{{\bf p}^\perp\cdot {\bf p}^{\pr \perp}}{m^2}}
{\sqrt{{\bf p}^2+m_e^2} - \sqrt{({\bf p}-{\bf k})^2+m_e^2}-|{\bf k}|}
\;.
\eea
Given the  constraint  $|{\bf k}| \leq \tilde{\lam} \ll |{\bf p}|$, this 
becomes
\bea
\del B_{L2}&\approx&~ 
\frac{\alpha}{4 \pi^2 }
 {\sum_{N^\pr}}^c \int \frac{d^3 k}{  |{\bf k}| }
~\theta\left(\tilde{\lam}-|{\bf k}|\right)
\int d^3 p \int d^3 p^\pr
\phi_N^\ast\left({\bf p}^\pr\right)
\phi_{N^\pr}\left({\bf p}^\pr\right)\nn\\
&&~~~~~~~~~~~~~~~~~~~~~~~~~~~~~~\times
\phi_{N^\pr}^\ast\left({\bf p}\right)\phi_N\left({\bf p}\right)
\frac{\frac{ {k^\perp}^2}{{k^+}^2}+\frac{ {k^\perp}^2}{k^+ m}+
\frac{{\bf p}^\perp\cdot {\bf p}^{\pr \perp}}{m^2}}
{|{\bf k}|}
\;.\label{eq:bl2}
\eea
This is the famous subtraction that Bethe performed in 1947 \cite{bethe}. 
In our approach it arose as a consequence of coupling coherence. 

$\del B_{L2}$ is infrared divergent ($k^+ \longrightarrow 0$)  as is
$\del B_{L1}$. This divergence
arises from the first two terms of $N_1$ (the ones independent
of ${\bf p}$ and ${\bf p}^\pr$). Noting that
\beaa
|{\bf k}| &=& \frac{1}{2} \left(\frac{{k^\perp}^2}{k^+}+k^+\right)
\stackrel{(k^+ \rightarrow 0)}{\longrightarrow}\frac{{k^\perp}^2}{2 k^+}
\;,
\eeaa
we have
\beaa
&&\beta_n-\beta_{n^\pr}-|{\bf k}| 
\stackrel{(k^+ \rightarrow 0)}{\longrightarrow}-\frac{{k^\perp}^2}{2 k^+}
\;,
\eeaa
and  these infrared divergent contributions from the 
first two terms of $N_1$  cancel, leaving an infrared finite shift,
\bea
\del B_L&=& \del B_{L1}+\del B_{L2}=\frac{\alpha}{4 \pi^2 }
 {\sum_{N^\pr}}^c \int \frac{d^3 k}{  |{\bf k}| }
~\theta\left(\tilde{\lam}-|{\bf k}|\right)
\int d^3 p \int d^3 p^\pr
\phi_N^\ast\left({\bf p}^\pr\right)
\phi_{N^\pr}\left({\bf p}^\pr\right)\nn\\
&&~~~~~~~~~~~~~~~~~~~~~~~~~~~~~~\times
\phi_{N^\pr}^\ast\left({\bf p}\right)\phi_N\left({\bf p}\right)
\frac{
{\bf p}^\perp\cdot {\bf p}^{\pr \perp}}{m^2}\left(\frac{1}
{\beta_n-\beta_{n^\pr}-|{\bf k}|}+\frac{1}{|{\bf k}|}\right)
\label{eq:bl1total}
\\
&=& \left(\frac{2}{3}\right)\frac{\alpha}{4 \pi^2 }
 {\sum_{N^\pr}}^c \int \frac{d^3 k}{  |{\bf k}| }
~\theta\left(\tilde{\lam}-|{\bf k}|\right)
\int d^3 p \int d^3 p^\pr
\phi_N^\ast\left({\bf p}^\pr\right)
\phi_{N^\pr}\left({\bf p}^\pr\right)\nn\\
&&~~~~~~~~~~~~~~~~~~~~~~~~~~~~~~\times
\phi_{N^\pr}^\ast\left({\bf p}\right)\phi_N\left({\bf p}\right)
\frac{
{\bf p}\cdot {\bf p}^{\pr }}{m^2}\left(\frac{1}
{\beta_n-\beta_{n^\pr}-|{\bf k}|}+\frac{1}{|{\bf k}|}\right)
\;.\label{eq:bl1total2}
\eea
This last step followed after averaging over directions 
 as dictated by 
rotational invariance. 

Eq.~(\ref{eq:bl1total2}) is easy to integrate, and our final result for the 
low-energy photon contribution is
\bea
\del B_L
&=&\frac{2 \alpha}{3 \pi m^2}
 {\sum_{N^\pr}}^c 
 \left(\beta_{n^\pr}-\beta_n\right) \log\left|
\frac{\tilde{\lam}+\beta_{n^\pr}-\beta_n}{\beta_{n^\pr}-\beta_n}\right|
\left|\langle
\phi_N|\hat{{\bf p}}|
\phi_{N^\pr}\rangle\right|^2
\label{eq:lastl1}\\
&=&\frac{2 \alpha}{3 \pi m^2}
 {\sum_{N^\pr}}^c 
 \left(\beta_{n^\pr}-\beta_n\right) \log\left|
\frac{\tilde{\lam}}{\beta_{n^\pr}-\beta_n}\right|
\left|\langle
\phi_N|\hat{{\bf p}}|
\phi_{N^\pr}\rangle\right|^2
\;,\label{eq:72}
\eea
where in this last step we recalled  $\tilde{\lam} \gg m \alpha^2$.
Note the $\tilde{\lam}$-dependence in the result.
This will cancel after we correctly add the contributions coming from high energy intermediate
photons,  which now follows.
\vskip.2in
\noindent
\centerline{{\bf B. High energy contribution}}
\vskip.2in

The high energy shift arises from three sources which are shown in Fig.~4. These are 
first-order BSPT shifts due to third and fourth order effective interactions
(see Appendix~B). 
The net result of these three diagrams is 
\bea
-\frac{\alpha}{2 \pi^2 {\bf q}^2} &\longrightarrow&
-\frac{\alpha}{2 \pi^2 {\bf q}^2}\left(
1+\del V_H
\right)\;,\label{eq:hbaby}
\eea
where ${\bf q}$ is the exchanged momentum of the electron, and
\bea
\del V_H&=&\del V_{H1}+\del V_{H2}+\del V_{H3}
\;,
\eea
with
\bea
\del V_{H1}&=&\frac{1}{2}
\int_k\theta\left({p_e^\pr}^+ -k^+\right)\theta\left(p_e^+ -k^+\right)
N_{H1} ~\theta_{H1}\nn\\
&&~~~~~~~~~~\times \left(
\frac{\left({\cal P}^+\right)^2}{\left(M_o^2-M^2\right)\left(M_o^2-{M^\pr}^2\right)}+
\frac{\left({\cal P}^+\right)^2}{\left({M_o^\pr}^2-M^2\right)\left({M_o^\pr}^2-{M^\pr}^2\right)}
\right)
\label{eq:vh1}\;,\\
\del V_{H2}&=&-\frac{1}{2}
\int_k\theta\left({p_e^\pr}^+ -k^+\right)\theta\left(p_e^+ -k^+\right)
N_{H2} ~\theta_{H2} ~
\frac{\left({\cal P}^+\right)^2}{\left(M_o^2-{M^\pr}^2\right)\left({M_o^\pr}^2-{M^\pr}^2\right)}
\label{eq:vh2}\;,\\
\del V_{H3}&=&-\frac{1}{2}
\int_k\theta\left({p_e^\pr}^+ -k^+\right)\theta\left(p_e^+ -k^+\right)
N_{H3}~ \theta_{H3} ~
\frac{\left({\cal P}^+\right)^2}{\left(M_o^2-{M}^2\right)\left({M_o^\pr}^2-{M}^2\right)}
\label{eq:vh3}
\;.
\eea
The factors  $\pm \frac{1}{2}$ in front arise from the form of the Bloch transformation
(see Eq.~(\ref{eq:heff}) of Appendix~B).
The vertex factors are given by
\bea
N_{H1}&=&\left(N_1\right)_{(k_1\rightarrow p_e^\pr-k,k_3 \rightarrow p_e-k)}
\label{eq:nh1}\;,\\
N_{H2}&=&\left(N_1\right)_{(k_1\rightarrow p_e^\pr-k,k_3 \rightarrow p_e^\pr-k,p_e\rightarrow
p_e^\pr)}
\label{eq:nh2}\;,\\
N_{H3}&=&\left(N_1\right)_{(k_1\rightarrow p_e-k,k_3 \rightarrow p_e-k,p_e^\pr\rightarrow
p_e)}
\label{eq:nh3}
\;,
\eea
 where $N_1$ was defined in Eq.~(\ref{eq:N1}). The free state masses are given by
 \bea
 M_o&=&\sqrt{{\bf p}^2+m_e^2}+\sqrt{{\bf p}^2+m_p^2}
 \;,\\
 M_o^\pr&=&\sqrt{{{\bf p}^\pr}^2+m_e^2}+\sqrt{{{\bf p}^\pr}^2+m_p^2}
 \;,\\
 M&=&|{\bf k}|+\sqrt{({\bf p}-{\bf k})^2+m_e^2}+\sqrt{{{\bf p}}^2+m_p^2}\;,
 \label{eq:ee}\\
 M^\pr&=&|{\bf k}|+\sqrt{({\bf p}^\pr-{\bf k})^2+m_e^2}+\sqrt{{{\bf p}^\pr}^2+m_p^2}\;.
 \label{eq:eee}
 \eea
 
 The Bloch transformation constrains the free masses of the states.
 As discussed before, the ``L" restrictions in Fig.~4 can be removed given
 $\tilde{\lam} \gg m \alpha^2$. However, the ``H" restrictions lead to important
 constraints given by the $\theta_H$ factors above, which we now discuss.
 They constrain  the free masses to satisfy (recall Eqs.~(\ref{eq:21})-(\ref{eq:21p})):
\bea
&\tilde{\lam}\leq&M-m_e-m_p \leq \tilde{\Lam}\;,\label{eq:85}\\
&\tilde{\lam}\leq&M^\pr-m_e-m_p \leq \tilde{\Lam}\label{eq:86}
\;,
\eea
where $M$ and $M^\pr$ are defined in Eqs.~(\ref{eq:ee}) and (\ref{eq:eee}) respectively.

As already mentioned, for convenience of calculation,
we will divide this high energy contribution into two
regions, $\tilde{\lam} \leq |{\bf k}| \leq b$ and $b \leq |{\bf k}| \leq m$ ({\em region one}
and {\em region two} respectively), with 
$m \alpha \ll b \sim m \sqrt{\alpha} \ll m$. Recall, $m \alpha^2 \ll \tilde{\lam}
 \sim m \alpha \sqrt{\alpha} \ll m \alpha$. We now show how this division
 into these two regions arises as a result of 
  the constraints of Eqs.~(\ref{eq:85}) and (\ref{eq:86}).
 
 In this first region, $|{\bf k}| \ll m$, and  Eqs.~(\ref{eq:85}) and (\ref{eq:86}) become
 \bea
 \tilde{\lam}&\leq& |{\bf k}| +\frac{({\bf p}-{\bf k})^2}{2 m}\sim ~|{\bf k}|~ \leq~ b~,\\
  \tilde{\lam}&\leq& |{\bf k}| +\frac{({\bf p}^\pr-{\bf k})^2}{2 m}\sim ~|{\bf k}|~ \leq ~b
 \;,
 \eea
 which is as we have already stated (recall that we always assume $m_p \rightarrow \infty$
 and drop the $1/m_p$ corrections since we are just after the dominant shift). 
 
  The analysis of the second region is slightly more complicated
 because $|{\bf k}| \gg m \alpha$, and near the upper limit $|{\bf k}| \sim m$. Since 
 $|{\bf k} |\gg m \alpha$ in this region, Eqs.~(\ref{eq:85}) and (\ref{eq:86}) both
 become
  \bea
 b&\leq& |{\bf k}|+\sqrt{{\bf k}^2+m^2}-m ~\leq~ \tilde{\Lam}
  \;.
 \eea
 This is just a linear constraint,  
 \bea
&& b\left(\frac{2m+b}{2m+2b}\right)~\leq~|{\bf k}|~\leq~ \frac{\tilde{\Lam}}{2}
\left(\frac{\tilde{\Lam}+2 m}{\tilde{\Lam}+m}\right)
 \;,
 \eea
 which, since we choose $b \ll m$, becomes
 \bea
 && b~\leq~|{\bf k}|~\leq~ \frac{\tilde{\Lam}}{2}
\left(\frac{\tilde{\Lam}+2 m}{\tilde{\Lam}+m}\right)
 \;.\label{eq:rrr}
 \eea
 The electron self-energy renormalization is performed in this paper, but
we do not deal with removing the full $\tilde{\Lam}$-dependence. 
 A full analysis of this dependence requires a complete 4th order calculation,
 which is beyond the scope of this paper. 
We cut off the photon momentum at the electron 
 mass, and proceed. 
 Note that from Eq.~(\ref{eq:rrr}), this choice corresponds to $\tilde{\Lam}^2=2 m^2$.
The point of calculating these high photon energy contributions 
is to  show that our results
 are independent of the effective cutoff, $\tilde{\lam}$.

 Taking a sample denominator we have
 \bea
 &&(M_o^2-M^2)=(M_o+M)(M_o-M)\approx
 2 m_p \left(\frac{{\bf p}^2}{2m}-|{\bf k}| -\frac{({\bf p}-{\bf k})^2}{2 m}\right)
 \approx ~-2 m_p|{\bf k}|
 \;,
 \eea
 in the {\em first} region; and
  \bea
 &&(M_o^2-M^2)=(M_o+M)(M_o-M)\approx 2 m_p \left(
 m-|{\bf k}|-\sqrt{{\bf k}^2+m^2}\right)\approx-2 m_p\left(|{\bf k}|+\frac{|{\bf k}|^2}{2m}\right)
 \;,
 \eea
 in the {\em second} region.
 
 Using these previous formulae, including
 ${\cal P}^+\longrightarrow m_p$ as $m_p \longrightarrow \infty$, Eqs.~(\ref{eq:vh1})-(\ref{eq:vh3}),
 after summing, become
 \bea
 \del V_H^\pr&=&
 -\frac{\alpha}{4 \pi} \frac{{q^\perp}^2}{m^2}
 \int_{-1}^{1} d \cos\theta \int_{\tilde{\lam}}^b \frac{d |{\bf k}|}
 {|{\bf k}|} \left(1 +
 {\cal O}\left(\frac{|{\bf k}|}{m}\right)
 \right)
 \;,
 \eea
 in the first region (the ``prime" on $\del V_H$ signifies the {\em first} region); and
 \bea
  \del V_H^{\pr \pr}&=&
  -\frac{\alpha}{4 \pi} \frac{{q^\perp}^2}{m^2}
  \int_{-1}^{1} d \cos\theta \int_{b}^m \frac{d |{\bf k}|}
 {|{\bf k}|} \left(1
 +c_n\frac{|{\bf k}|}{m}
 \left(1+\cos\theta\right)-c_d\frac{|{\bf k}|}{m}
 \right.\nn \\
 &&~~~~~~~~~~~~~~~~~~~~~~~~~~~~~~~~~~~~~~~~+\left.
 {\cal O}\left(\frac{|{\bf k}|^2}{m^2}\right)
 \right)
  \;,
 \eea
 in the second region (the ``double prime" on $\del V_H$ signifies the {\em second} region). 
 In the second region since the photon momentum is not necessarily smaller
 than the electron mass, we have kept two terms in the $|{\bf k}|/m$ expansion of the integrand.
 In the ${\cal O}(|{\bf k}|/m)$ term we have introduced two constants, $c_n$ and $c_d$, which
 denote {\em numerator} and energy {\em denominator} corrections respectively. Hereafter we
 set $c_n=1$ and $c_d=1$, as given by the theory.
 Note that in combining $\del V_{H1}$, $\del V_{H2}$ and $\del V_{H3}$, many cancelations
 occur; most noteworthy, 
  each contribution is individually infrared divergent ($k^+ \rightarrow 0$), but in the sum
 the divergences cancel.
 These final equations are easily integrated, and we have
 \bea
 \del V_H^\pr&=&
 -\frac{\alpha}{2 \pi} \frac{{q^\perp}^2}{m^2}\log\left(\frac{b}{\tilde{\lam}}\right)
 \;,\label{eq:1}\\
  \del V_H^{\pr \pr}&=&
  -\frac{\alpha}{2 \pi} \frac{{q^\perp}^2}{m^2}\log\left(\frac{m}{b}\right)
  \;.\label{eq:2}
 \eea
 In the second region note that the  ${\cal O}(|{\bf k}|/m)$ terms coming
 from numerator and energy denominator corrections 
  cancel, leaving the ${\cal O}(1)$ piece alone.
 The combined high energy contribution is
 \bea
&& \del V_H=\del V_H^\pr+\del V_H^{\pr \pr}=
-\frac{\alpha}{2 \pi} \frac{{q^\perp}^2}{m^2}\log\left(\frac{m}{\tilde{\lam}}\right)
\label{eq:highfinal}
 \;,
 \eea
 which is independent of $b$,
 as required for consistency. Recall that 
 ${\bf q}={\bf p}^\pr-{\bf p}$: the difference between the final and initial
 electron momenta.
 
 From the definition of $\del V_H$
 (see Eq.~(\ref{eq:hbaby})), we see that
 this correction shifts the energy levels an amount
 \bea
 \del B_H&=&-\frac{\alpha}{2 \pi^2}\int d^3p ~
 d^3 p^\pr \phi_N^\ast\left({\bf p}^\pr\right) \left(
 \frac{\del V_H}{\pp}
 \right)\phi_N\left({\bf p}\right)
 \;.
 \eea
 Combining this with Eq.~(\ref{eq:highfinal}) gives
\bea
 \del B_H&=&\frac{\alpha^2}{4 \pi^3 m^2}
\log\left(\frac{m}{\tilde{\lam}}\right)\int d^3p ~
 d^3 p^\pr \phi_N^\ast\left({\bf p}^\pr\right) \left(
 \frac{\left({\bf p}^\perp-{\bf p}^{\pr \perp}\right)^2}{\pp}
\right)\phi_N\left( {\bf p} \right) \label{eq:100}\\
 &=&\frac{\alpha^2}{6 \pi^3 m^2}
 \log\left(\frac{m}{\tilde{\lam}}\right)\left(\int d^3p ~
 \phi_N\left({\bf p}\right)\right)^2
 \;,\label{eq:101}
 \eea
 where in this last step we averaged over directions and noted that the wave function
 at the origin is real. For more details on this averaging over directions see
 Appendix~C.
\vskip.2in
\noindent
\centerline{{\bf C. Total contribution}}
\vskip.2in

In this section we combine the results of the last two sections for the low and
high photon energy contributions, and perform the required sums/integrations
to calculate the total shift between the $2S_{\frac{1}{2}}$ and $2P_{\frac{1}{2}}$ energy levels
of hydrogen.

Adding Eqs.~(\ref{eq:72}) and (\ref{eq:101}) gives for the total shift
\bea
\del B&=&\del B_L+\del B_H=\frac{2 \alpha}{3 \pi m^2}
 {\sum_{N^\pr}}^c 
 \left(\beta_{n^\pr}-\beta_n\right) \log\left|
\frac{\tilde{\lam}}{\beta_{n^\pr}-\beta_n}\right|
\left|\langle
\phi_N|\hat{{\bf p}}|
\phi_{N^\pr}\rangle\right|^2\nn\\
&&~~~~~~~~~~~~~~~~~~~~~~~~~~+~\frac{\alpha^2}{6 \pi^3 m^2}
 \log\left(\frac{m}{\tilde{\lam}}\right)\left(\int d^3p ~
 \phi_N\left({\bf p}\right)\right)^2
\;.\label{eq:blog}\eea

For the second term we have
\bea
\left(\int d^3 p~ \phi_N\left({\bf p}\right)\right)^2
 &=&  \left((2 \pi)^{\frac{3}{2}}\phi_N\left({\bf x}=0\right)\right)^2
=\frac{(2 \pi)^{3}} {\pi}\left(\frac{m \alpha}{n}\right)^3\del_{l,0}
\;.\label{eq:103}
\eea
The $(2 \pi)^3$ factor arose because of our normalization choice (see Eq.~(\ref{eq:norm})).

The first term of Eq.~(\ref{eq:blog}) 
is the famous Bethe log and must be calculated numerically, summing
over all bound and continuum states. Following standard convention we define
an average excitation energy, $\overline{\beta}(n,l)$, 
\bea
\log\left(\frac{\overline{\beta}(n,l)}{Ryd}\right)
{\sum_{N^\pr}}^c 
 \left(\beta_{n^\pr}-\beta_n\right) 
\left|\langle
\phi_N|\hat{{\bf p}}|
\phi_{N^\pr}\rangle\right|^2
=
{\sum_{N^\pr}}^c 
 \left(\beta_{n^\pr}-\beta_n\right) \log\left|
\frac{\beta_{n^\pr}-\beta_n}{Ryd}\right|
\left|\langle
\phi_N|\hat{{\bf p}}|
\phi_{N^\pr}\rangle\right|^2
\;.\label{eq:44}\eea
The sum on the left is evaluated by standard techniques ($H_c=p^2/(2m)-\alpha/r$ ),
\bea
&&{\sum_{N^\pr}}^c 
 \left(\beta_{n^\pr}-\beta_n\right) 
\left|\langle
\phi_N|\hp|
\phi_{N^\pr}\rangle\right|^2
=\frac{1}{2}\langle
\phi_N\left|
\left[\hp,H_c\right]\cdot \hp+\hat{{\bf p}}
\cdot\left[H_c,\hp\right]\right|
\phi_{N}\rangle\nn\\
&&~~~~~~~~~~=-\frac{1}{2}\langle
\phi_N\left|\left[
\hp
\cdot,\left[\hp,H_c\right]\right]\right|
\phi_{N}\rangle=-\frac{1}{2}\left\langle
\phi_N\left|\left[
\hp
\cdot,-i\vec{\nabla}\left(-\alpha/r\right)\right]\right|
\phi_{N}\right\rangle\nn\\
&&~~~~~~~~~~=
-\frac{1}{2}\left\langle
\phi_N\left|(-i)^2{\vec{\nabla}}^2\left(-\alpha/r\right)\right|
\phi_{N}\right\rangle=
-\frac{1}{2}(-i)^2 (-\alpha) (-4\pi)\left\langle
\phi_N\left|\del^3\left(r\right)\right|
\phi_{N}\right\rangle\nn\\
&&~~~~~~~~~~=2  \alpha \left(\frac{m \alpha}{n}\right)^3\del_{l,0}\;.
\eea
This vanishes for $l \neq 0$, but the average excitation energy, $\overline{\beta}(n,l)$,
is  defined (it is just a way to catalogue the numerical sum on the right
of Eq.~(\ref{eq:44}), the quantity
we need to know)
with the sum on the
left hand side set to its value for $l=0$. In summary, $\overline{\beta}(n,l)$
for all states is defined by
\bea
\log\left(\frac{\overline{\beta}(n,l)}{Ryd}\right)
2  \alpha \left(\frac{m \alpha}{n}\right)^3
=
{\sum_{N^\pr}}^c 
 \left(\beta_{n^\pr}-\beta_n\right) \log\left|
\frac{\beta_{n^\pr}-\beta_n}{Ryd}\right|
\left|\langle
\phi_N|\hat{{\bf p}}|
\phi_{N^\pr}\rangle\right|^2
\;.\label{eq:45}\eea
Without further ado, this sum has been evaluated  by R. W. Huff \cite{huff}.
His results for the $n=2$ levels are
\bea
\overline{\beta}(2,0)&=&
16.63934203(1)~
Ryd\;,\label{eq:107}\\
\overline{\beta}(2,1)&=&
0.9704293186(3)~
Ryd
\;,\label{eq:108}
\eea
where the figures in parentheses give the
number of units of estimated error in the last decimal place (R. W. Huff's estimates).

Combining the results:
\bea
\del B_{2S_{\frac{1}{2}}}&=&\frac{2 \alpha}{3 \pi m^2} 
\log\left(\frac{\tilde{\lam}}{\overline{\beta}(2,0)}\right)
2  \alpha \left(\frac{m \alpha}{n}\right)^3+\frac{\alpha^2}{6 \pi^3 m^2} 
\log\left(\frac{m}{\tilde{\lam}}\right)
\frac{(2 \pi)^{3}} {\pi}\left(\frac{m \alpha}{n}\right)^3\nn\\
&=& \frac{ \alpha^3 Ryd}{3 \pi }
\log\left(\frac{m}{\overline{\beta}(2,0)}\right)\;,\\
\del B_{2P_{\frac{1}{2}}}&=&\frac{2 \alpha}{3 \pi m^2} 
\log\left(\frac{Ryd}{\overline{\beta}(2,1)}\right)
2  \alpha \left(\frac{m \alpha}{n}\right)^3\nn\\
&=& \frac{ \alpha^3 Ryd}{3 \pi }
\log\left(\frac{Ryd}{\overline{\beta}(2,1)}\right)
\;.
\eea
Note the cancelation of the $\tilde{\lam}$-dependence.
Thus, the Lamb shift is 
\bea
\del B_{_{Lamb}}&=&\del B_{2S_{\frac{1}{2}}}-\del B_{2P_{\frac{1}{2}}}=
\frac{ \alpha^3 Ryd}{3 \pi }
\log\left(\frac{m~\overline{\beta}(2,1)}{Ryd~\overline{\beta}(2,0)}\right)
\\
&=& (1047-4)~{\rm MHz}~ (2 \pi \hbar)= 1043~{\rm MHz}~ (2 \pi \hbar)
\;, 
\eea
where we use Ref.~\cite{expt} and
the average excitation energies of Eqs.~(\ref{eq:107}) and (\ref{eq:108}).
Note that the $2P_{\frac{1}{2}}$ shift is only about one half of a percent of the
 $2S_{\frac{1}{2}}$ shift.
\vskip.2in
\noindent
\centerline{{\bf V. SUMMARY AND DISCUSSION}}
\vskip.2in

In a light-front Hamiltonian approach,
 we have shown how to do a consistent Lamb shift calculation for the $n=2$, $j=1/2$
levels of hydrogen over the photon energy scales
\beaa
&&0\lr m \alpha^2\lr\tilde{\lam}\lr m \alpha \lr b \lr m
\;,
\eeaa
with the choices $m \alpha^2 \ll \tilde{\lam} \ll m \alpha$ and $m \alpha \ll b \ll m$.
In a consistent set of diagrams we showed how $\tilde{\lam}$- and $b$-dependence cancel
leaving the dominant part of the Lamb shift,
 $1043$ MHz. For completeness, the 
  $n=2$ spectrum of hydrogen is shown in Fig.~5.
 
 If we
 compare the three regions we see the following results (we only compare for the
 $2S_{\frac{1}{2}}$ shift since the $2P_{\frac{1}{2}}$ shift is negligible
 within our errors):
\bea
&&(0\leq|{\bf k}|\leq\tilde{\lam})~~~~~
\del B_{_{Lamb}}\sim
\frac{ \alpha^3 Ryd}{3 \pi }
\log\left(\frac{\tilde{\lam}}{16.64 ~Ryd}\right)\sim
 46~{\rm MHz}~ (2 \pi \hbar)\sim 4\%\;,
 \label{eq:ttbbg}\\ 
&&(\tilde{\lam}\leq|{\bf k}|\leq b)~~~~~
\del B_{_{Lamb}}\sim 
\frac{ \alpha^3 Ryd}{3 \pi }
\log\left(\frac{b}{\tilde{\lam}}\right)\sim
667~{\rm MHz} ~(2 \pi \hbar)\sim 64\%\;,\\
&&(b\leq|{\bf k}|\leq m)~~~~~\del B_{_{Lamb}}\sim 
\frac{ \alpha^3 Ryd}{3 \pi }
\log\left(\frac{m}{b}\right)\sim
334~{\rm MHz}~ (2 \pi \hbar)\sim 32\%
\;,
\eea
where we used $\tilde{\lam} = m \alpha \sqrt{\alpha}$ and $b= m \sqrt{\alpha}$, consistent
choices used throughout 
 this paper.
As expected on physical grounds (see the Introduction),  photons
with momentum
\bea
&&|{\bf k}| \sim 1/a_o \sim m \alpha\;,
\eea
 couple the strongest to the hydrogen system. As seen above,  the effects
of photons of this
momentum amounted to about $2/3$ of the Lamb shift, the dominant part of this experimentally
observed shift.

In this paper,  the one loop electron self-energy renormalization was performed.
The complete one loop renormalization was not needed to obtain the dominant
part of the Lamb shift. Our answer, $1043$ MHz, turned out to be accurate. 
However, to obtain more precision, the full one loop renormalization must be performed
of course.
Also, each of our five diagrams (of Figures~2 and 4) were infrared ($k^+_{photon} \longrightarrow 0$)
divergent. However, both the sum of the two low photon
energy diagrams and the sum of the 
three high photon energy diagrams were infrared finite.

The state of the art for the bound-state problem in a light-front Hamiltonian
gauge theory in four dimensions has been advanced in this paper.
In applying these methods to QCD, the general computational strategy described in the Introduction
 does not change. However, gluons carry color charge and interact
strongly at low energies, thus the answers to the three questions posed in the initial paragraph
of the Introduction change drastically.
For a constituent picture to emerge the massless gluons must be confined,
so that it costs energy to add a low momentum gluon to the system.\footnote{The energy of
this confined low momentum gluon can be interpreted as an effective gluon mass
if it is convenient.}
It has been shown that the second order effective interactions (including the very important
first order instantaneous-gluon potential) are confining \cite{perrybrazil}, which is 
promising.
Given confinement, we can lower the effective cutoff below the gluon production 
threshold perturbatively and obtain a constituent approximation.
As in QED, we can not lower the effective cutoff
below the non-perturbative bound-state energy scale.
Thus Eq.~(\ref{eq:2cv}) of hydrogen in QCD becomes
\bea
&&\Lam_{QCD} \ll \tilde{\lam} \ll E_{gluon} \sim M_{glueball}/2
~.\eea
Since $\Lam_{QCD}$ ranges from $200$-$400$ MeV (depending on the
renormalization scheme that is chosen)
and $M_{glueball}$ ranges from  $1500$-$1700$ MeV, this constraint can be
satisfied and it becomes plausible
to attack QCD by the same computational strategy that was outlined and carried
out for the Lamb shift in QED in this paper.
\newpage
\noindent
\centerline{{\bf  ACKNOWLEDGMENTS}}
\vskip.2in
The authors 
 wish to acknowledge useful discussions with Brent H. Allen, Martina M. Brisudov\'{a},
Stanis{\l}aw D. 
 G{\l}azek, Koji Harada and Kenneth G. Wilson.
Research reported in this paper has been supported  by 
the National Science Foundation under grant PHY-9409042. 
\vskip.2in
\noindent
\centerline{{\bf  APPENDIX A: LIGHT FRONT CONVENTIONS}}
\vskip.2in

In this Appendix we  write our light-front conventions for the electron, proton, and photon
system.
\beaa
&\bullet&  V^{\pm}=V^0 \pm V^3~~\rm{where}~V^\mu ~{\rm is~ any~ four~vector.}\\
&\bullet& 
\gamma^+=\left[
\begin{array}{cc}
0&0\\
2 i&0
\end{array}
\right]~;~
\gamma^-=\left[
\begin{array}{cc}
0&-2 i\\
0&0
\end{array}
\right]\\
&\bullet&
\alpha^i=\gamma^0 \gamma^i=\left[
\begin{array}{cc}
0&\sig^i\\
\sig^i&0
\end{array}
\right]~;~i=1,2~;~\sig^i~{\rm are}~SU(2)~{\rm Pauli}~{\rm matrices.}\\
&\bullet&\Lambda_+=\frac{1}{2}\gamma^0\gamma^+=
\left[
\begin{array}{cc}
1&0\\
0&0
\end{array}
\right]~;~
\Lambda_-=\frac{1}{2}\gamma^0\gamma^-=
\left[
\begin{array}{cc}
0&0\\
0&1
\end{array}
\right]\\
&\bullet&\psi_\pm=\Lambda_\pm \psi~;~\psi=\psi_+ + \psi_-\\
&\bullet&\psi_{e+}=\left[
\begin{array}{c}
\xi_e\\
0
\end{array}
\right]~;~
\psi_{p+}=\left[
\begin{array}{c}
\xi_p\\
0
\end{array}
\right]~;~ e~{\rm for~ electron}, ~p~{\rm for~ proton.}\\
&\bullet&\psi_{e-}=\left[
\begin{array}{c}
0\\
\frac{1}{i \partial^+}\left[\left(
\sig^i\left(
i \partial^i+e A^i
\right)+i m_e
\right)\xi_e\right]
\end{array}
\right]\\
&\bullet&\psi_{p-}=\left[
\begin{array}{c}
0\\
\frac{1}{i \partial^+}\left[\left(
\sig^i\left(
i \partial^i-e A^i
\right)+i m_p
\right)\xi_p\right]
\end{array}
\right]\\
&\bullet&
A^-= \frac{-2}{( \partial^+)^2} J^+ + 2 \frac{ \partial^i}{ \partial^+} A^i\\
&\bullet& J^+=2 e \left(
\xi_p^\dagger \xi_p-\xi_e^\dagger \xi_e
\right)~;~ e > 0\\
&\bullet&A^+=0
\eeaa

In momentum space the field
operators are expanded as (at $x^+ = 0$):
\beaa
A^i(x)&=&\sum_{s = \pm 1} \int_{q} (\eps^i_s a_s(q)
e^{-iq \cdot x}+h.c.)\;,\\
\xi_e(x)&=&\sum_{s = \pm 1}\chi_s \int_p \sqrt{p^+}(b_s(p)
e^{-ip \cdot x}+d_{{\overline s}}^{\dagger}(p)e^{+ip\cdot x})\;,\\
\xi_p(x)&=&\sum_{s = \pm 1}\chi_s \int_p \sqrt{p^+}(B_s(p)
e^{-ip \cdot x}+D_{{\overline s}}^{\dagger}(p)e^{+ip\cdot x})\;,\\
{\rm with}
&&\eps_1^i=\frac{-1}{\sqrt{2}}\left(\delta_{i,1} + i ~\delta_{i,2}\right)
\;,\;\eps_{-1}^i=\frac{1}{\sqrt{2}}\left(
\delta_{i,1} - i~ \delta_{i,2}\right)\;,\\
&&\chi_{_{1}}=
\left(\begin{array}{c}
1\\
0
\end{array}\right)\;,\;\chi_{_{\overline 1}}=
\left(\begin{array}{c}
0\\
1
\end{array}\right)
\;.
\eeaa
Note that ${\overline s} \equiv -s$. Also, we are using the shorthand
\beaa
\int_p f(p)&=&\int\frac{d^4p}{(2 \pi)^4}~2 \pi~ \del (p^2-m^2)~ \theta(p^0)~f(p)=
\int\frac{d^2p^\perp dp^+ \theta(p^+)}{16 \pi^3 p^+}f(p)
{\Biggr{|}}_{p^-=\frac{{p^\perp}^2+m^2}{p^+}}
~.
\eeaa
The fermion  helicity can only take
on the  values
$ \pm 1/2$, however we define: 
$h_3=  s/2$; therefore, ``s" can only take
on the values
$\pm 1$. 
The commutation (anti-commutation) relations and free Fock states are given by:
\beaa
&&[a_\lambda(q),a_{\lambda^\prime}^{\dagger}(q^\prime)]=16\pi^3q^+
\del^3(q-q^\prime)\del_{\lambda \lambda^\prime}\;,\;(\;\del^3(p)=
\del^2(p^\perp)\del(p^+)\;)\;,\\
&&\{b_s(p),b_{s^\prime}^{\dagger}(p^\prime)\}=
\{d_s(p),d_{s^\prime}^{\dagger}(p^\prime)\}=
16\pi^3p^+
\del^3(p-p^\prime)\del_{s s^\prime}\;,\\
&&\{B_s(p),B_{s^\prime}^{\dagger}(p^\prime)\}=
\{D_s(p),D_{s^\prime}^{\dagger}(p^\prime)\}=
16\pi^3p^+
\del^3(p-p^\prime)\del_{s s^\prime}\;,\\
&&\langle p_1 s_1 | p_2 s_2 \rangle = 16\pi^3p_1^+
\del^3(p_1-p_2)\del_{s_1 s_2}\;,\;|p_1 s_1\rangle=
b_{s_1}^{\dagger}(p_1)|0\rangle\;,\;etc\;.
\eeaa
\vskip.2in
\noindent
\centerline{{\bf  APPENDIX B: BLOCH TRANSFORMATION}}
\vskip.2in

In this Appendix we discuss a derivation of our effective Hamiltonian
via a Bloch transformation \cite{bloch}.
We use the Bloch transformation to separate the low and high energy scales 
of the problem
and derive an effective Hamiltonian acting in the low energy space alone with an identical
  low energy spectrum
 to the bare Hamiltonian. In this Appendix,
we closely follow    \S IV of Ref.~\cite{whitebook}, where
a  discussion, including the original references, and
derivation of  a general 
effective Bloch Hamiltonian  can be found. 

We start with a bare time-independent Schr{\"o}dinger equation: 
\bea
&&H_\Lam
 |\Psi_\Lam\rangle=E |\Psi_\Lam\rangle
\;.\label{eq:eeeee}
\eea
Then
projection operators onto the low and high energy spaces, $P_L$ and 
$P_H$ respectively,
are defined,

\bea
P_L&=&\theta\left(
\frac{\lam^2+{{\cal P}^\perp}^2}{{\cal P}^+}-h
\right)
,\\
P_H&=&
\theta\left(
\frac{\Lam^2+{{\cal P}^\perp}^2}{{\cal P}^+}-h
\right)~\theta\left(h-
\frac{\lam^2+{{\cal P}^\perp}^2}{{\cal P}^+}
\right)
\;,\\
\tl+\th&=&\theta\left(
\frac{\Lam^2+{{\cal P}^\perp}^2}{{\cal P}^+}-h
\right)
\;,
\eea
where $\theta(x)$ is a step function. Then
  an effective Hamiltonian acting in the low energy space alone with an equivalent 
low energy spectrum to $H_\Lam$ is sought. 
$\Lam$ and $\lam$ are the bare and effective cutoffs respectively with
 $\lam < \Lam$.\footnote{The shorthands 
 $\frac{\lam^2+{{\cal P}^\perp}^2}{{\cal P}^+} \longrightarrow \lam$ and 
 $\frac{\Lam^2+{{\cal P}^\perp}^2}{{\cal P}^+}\longrightarrow \Lam$ are often used when
 it does not lead to confusion.}  ${\cal P}=\left({\cal P}^+,{\cal P}^\perp\right)$ is the total momentum
 of the hydrogen state.
 $h$ is the free Hamiltonian of the hydrogen system of Eq.~(\ref{eq:freebaby}).

Proceeding, a new operator, ${\cal R}$, is defined
that connects the $\tl$ and $\th$ spaces:
\bea
&&\th |\Psi_\Lam\rangle={\cal R} \tl |\Psi_\Lam\rangle
\;.
\label{eq:RRR}
\eea
More explicitly ($\sum_n |n\rangle\langle n|=1$):
\bea
{\cal R}&=&\sum_{n,m}\langle n|\Psi_\Lam\rangle\langle\Psi_\Lam|m\rangle
\frac{ \th |n\rangle\langle m|\tl}
{\langle\Psi_\Lam|\tl|\Psi_\Lam\rangle}
\;,
\eea
which shows that ${\cal R} \th|\Psi_\Lam\rangle=0$ and ${\cal R}^\dag \tl|\Psi_\Lam
 \rangle=0$, etc.
For the construction of ${\cal R}$, see below Eq.~(4.4) in Ref.~\cite{whitebook}.

This leads to the following 
 time-independent Schr{\"o}dinger equation for the effective Hamiltonian 
\bea
&&H_\lam |\Phi_\lam\rangle=E |\Phi_\lam\rangle
\;.
\eea
$E$ is the same eigenvalue as in Eq.~(\ref{eq:eeeee}). The state $|\Phi_\lam\rangle$ is a  projection
onto the low energy space with the same norm as $|\Psi_\Lam\rangle$ in Eq.~(\ref{eq:eeeee}):
\bea
|\Phi_\lam\rangle&=&\sqrt{1+{\cal R}^\dag {\cal R}}~\tl|\Psi_\Lam\rangle
\;;
\label{eq:normmm}
\eea
 $H_\lam$ is a hermitian effective Hamiltonian given by
\bea
H_\lam=\frac{1}{\sqrt{1+{\cal R}^\dag {\cal R}}}
\left(\tl+{\cal R}^\dag\right)
H_\Lam
\left(\tl+{\cal R}\right)
\frac{1}{\sqrt{1+{\cal R}^\dag {\cal R}}}
\;.
\label{eq:ttrr}
\eea
Note that $H_\lam$ acts in the low energy space alone. 
To summarize,   $H_\lam$ of Eq.~(\ref{eq:ttrr})  is guaranteed to have
the same low energy spectrum as the bare Hamiltonian, $H_\Lam$; also, after diagonalizing
$H_\lam$, the bare state, $|\Psi_\Lam\rangle$, of Eq.~(\ref{eq:eeeee}), if desired,
 is obtained through Eqs.(\ref{eq:normmm}) and
(\ref{eq:RRR}).\footnote{This state, $|\Psi_\Lam\rangle$,
 will span the whole space, but will correspond to the respective
low energy eigenvalue. The bare states that correspond to the respective
high energy eigenvalues can not be obtained from the effective Hamiltonian, 
$H_\lam$; we must of course use the bare Hamiltonian, $H_\Lam$, to accomplish
this task.}
 
 Defining $H_\Lam=h+v_\Lam$, where $h$ is the free field theoretic Hamiltonian and
 $v_\Lam$ are the bare interactions,\footnote{$h$ is written in terms of renormalized parameters,
 and it is convenient to define $v_\Lam = v + \del v_\Lam$, where $v$ is the canonical field
 theoretic interactions written in terms of renormalized parameters
  and $\del v_\Lam$ are  the counterterms
 that must be determined through the process of renormalization. See  Eqs.~(\ref{eq:can1})
 and (\ref{eq:can2})
 for the canonical Hamiltonian of the hydrogen system.}
  to third order in $v_\Lam$, the effective Hamiltonian is given by
 \bea
 \langle a | H_\lam |b \rangle&=&\langle a|h+v_\Lam|b\rangle
 +\frac{1}{2}\sum_i\left(
\frac{ \langle a | v_\Lam|i\rangle\langle i|v_\Lam|b\rangle}{\Delta_{ai}}+
 \frac{\langle a | v_\Lam|i\rangle\langle i|v_\Lam|b\rangle}{\Delta_{bi}}
 \right)\nn\\
 &&~+\frac{1}{2}\sum_{i,j}\left(
 \frac{\langle a | v_\Lam|i\rangle\langle i|v_\Lam|j\rangle\langle j|v_\Lam|b\rangle }{\Del_{ai}\Del_{aj}}+
 \frac{\langle a | v_\Lam|i\rangle\langle i|v_\Lam|j\rangle\langle j|v_\Lam|b\rangle}{\Del_{bi}\Del_{bj}}
 \right)\nonumber\\
 &&~-\frac{1}{2}\sum_{c,i}\left(
 \frac{\langle a | v_\Lam|i\rangle\langle i|v_\Lam|c\rangle\langle c|v_\Lam|b\rangle }{\Del_{bi}\Del_{ci}}+
 \frac{\langle a | v_\Lam|c\rangle\langle c|v_\Lam|i\rangle\langle i|v_\Lam|b\rangle}{\Del_{ai}\Del_{ci}}
 \right)\label{eq:heff}\\
 &&~+~{\cal O}\left(\sum \langle v_\Lam\rangle^4/(\veps_{high}-\veps_{low})^3\right)
 \;.\nn
 \eea
 $\Delta_{ia}= \veps_i-\veps_a$, with $h |i\rangle = \veps_i |i\rangle$.
 We are using $|a\rangle,~|b\rangle,~\cdots$ 
 to denote low energy states (states in $\tl$) and $|i\rangle,~|j\rangle,~\cdots$
 to denote high energy states (states in $\th$). See the already 
 mentioned Ref.~\cite{whitebook} for a description of an arbitrary order (in perturbation theory)
 effective Hamiltonian
 and also for a convenient diagrammatic representation of the same.
 \newpage

\noindent
\centerline{{\bf  APPENDIX C: AVERAGING OVER DIRECTIONS}}
\vskip.2in

In this Appendix we calculate the following Coulomb matrix element:
\bea
I_\perp&=&\int d^3 p \int d^3 p^\pr\;\phi_N^{ \ast}({\bf p}^\pr) \frac
{\left({\bf p}^\perp-{\bf p}^{\pr \perp}\right)^2}{\pp}\phi_{N}({\bf p})
\;,
\eea
to verify the step taken from Eq~(\ref{eq:100}) to (\ref{eq:101}) in the paper.
It is useful to define another integral,
\bea
I_z&=&\int d^3 p \int d^3 p^\pr\;\phi_N^{ \ast}({\bf p}^\pr) \frac
{\left({\bf p}_z-{\bf p}^\pr_z\right)^2}{\pp}\phi_{N}({\bf p})
\;.
\eea
Now note that
\bea
I&=&I_\perp+I_z=\int d^3 p \int d^3 p^\pr\;\phi_N^{ \ast}({\bf p}^\pr) 
\phi_{N}({\bf p})
=\frac{(2 \pi)^{3}} {\pi}\left(\frac{m \alpha}{n}\right)^3\del_{l,0}
\;,\label{eq:yyyy}
\eea
where in this last step we recalled Eq.~(\ref{eq:103}) and the fact that
the wave function at the origin is real.

For $l=0$, the wave function satisfies
\bea
&&\phi_{n,0,0}({\bf p})=\phi_{n,0,0}(|{\bf p}|)
\;.
\eea
Thus, by symmetry, for $l=0$,
\bea
&&I_z=\frac{1}{3} I=\frac{(2 \pi)^{3}} {3\pi}\left(\frac{m \alpha}{n}\right)^3
\;,
\eea
which from Eq.~(\ref{eq:yyyy}), for $l=0$, gives
\bea
&&I_\perp=\frac{2}{3} I=\frac{2(2 \pi)^{3}} {3\pi}\left(\frac{m \alpha}{n}\right)^3
\;,
\eea

For $l \neq 0$, first note that $I=0$. Thus, for $l \neq 0$,
\bea
I_\perp=-I_z
\;;\label{eq:131}
\eea
we will calculate $I_z$ below which then implies $I_\perp$.
Next
note that  in position space 
\bea
I&=&-2 \pi^2\int d^3 x~ \phi_N^\ast\left({\bf x}\right)\left(
{\vec{\nabla}}^2\frac{1}{|{\bf x}|}
\right) \phi_N\left({\bf x}\right)
\;,
\eea
using ${\vec{\nabla}}^2\frac{1}{|{\bf x}|}=-4 \pi \del^3 \left({\bf x}\right)$.
 Thus, for $l \neq 0$,
 in position space
\bea
I_z&=&-2 \pi^2\int d^3 x ~\phi_N^\ast\left({\bf x}\right)\left(
{\vec{\nabla}}_z^2\frac{1}{|{\bf x}|}
\right) \phi_N\left({\bf x}\right)
\;.
\eea
Note that there is no $|{\bf x}|\rightarrow 0$ ambiguity in this previous
equation because for $l \neq 0$, the wave function vanishes at the origin.
Carrying out the derivative gives
\bea
I_z&=&-2 \pi^2\int d^3 x ~\phi_N^\ast\left({\bf x}\right)\left(
\frac{-1+3z^2/|{\bf x}|^2}{|{\bf x}|^3}
\right) \phi_N\left({\bf x}\right)
\;.\label{eq:lll}
\eea
This matrix element was performed in the first appendix of
Bethe and Salpeter's textbook \cite{bethetext}. We use two of 
their formulas, (3.26) and (A.29).\footnote{Warning to the reader: In this text, they use atomic units,
$\hbar=c=m=\alpha=1$, so $m$ and $\alpha$ have to be placed back into the formulas.}
Eq.~(\ref{eq:lll}) integrated gives
\bea
I_z&=&-2 \pi^2 ~\overline{r^{-3}}~ c(l,m_l)
\;,
\eea
with
\bea
\overline{r^{-3}}&=&\frac{1}{l(l+1)(l+\frac{1}{2})} \left(\frac{m \alpha}{n}\right)^3\;,
\\
c(l,m_l)&=&-1+3 \left(
\frac{2 l^2+2 l-1-2 m_l^2}
{(2 l+3)(2 l-1)}
\right)
\;.
\eea
Thus, recalling Eq.~(\ref{eq:131}), our result for $l\neq 0$ is
\bea
I_\perp&=&2 \pi^2 ~\overline{r^{-3}}~ c(l,m_l)
\;.\label{eq:nhy}
\eea

For $l=1$, $I_\perp$ is not zero, so what is going on?  The answer lies in the fact 
that we really want to take matrix elements in the $|j,m_j\rangle$ basis not the 
$|m_l,m_s\rangle$ basis, and based on rotational invariance,
our results should be independent of $m_j$. 
To proceed, note that the interactions we considered in this paper
conserved $m_s$ and  our matrix elements were independent of $m_s$. 
Next, note that the result,  Eq.~(\ref{eq:nhy}), is even under $m_l \longrightarrow
-m_l$. Given this, the Clebsch-Gordan coefficients for the
$2 P_{\frac{1}{2}}$ states imply
\bea
\langle j=1/2,m_j|V|j=1/2,m_j\rangle&=&\frac{1}{2l+1}\sum_{m_l=-l}^{l}
\langle m_l|V|m_l\rangle
\;,
\eea
where $l=1$.
Now note that $I_\perp$ given by Eq.~(\ref{eq:nhy}) averaged over $m_l$ vanishes,
\bea
\frac{1}{2l+1}\sum_{m_l=-l}^{l}I_\perp=0
\;,\label{eq:GGG}
\eea
where  we used
\bea
\frac{1}{2l+1}\sum_{m_l=-l}^{l}m_l^2=\frac{1}{3}l(l+1)
\;,
\eea
an obvious result after the answer is known.
This result (Eq.(\ref{eq:GGG})) was used in the step that led from Eq.~(\ref{eq:100}) 
to Eq.~(\ref{eq:101}) in the paper, and this Appendix is now complete.

\underline{Figure captions}

Figure 1:  The effective interactions that add to give the Coulomb potential. ``H"
implies that the photon energy is greater than $\tilde{\lam}$. ``L" implies that the
electron kinetic energy is less than $\tilde{\lam}$. We choose $m \alpha^2 \ll \tilde{\lam}
\ll m \alpha$; these ``H" and ``L" constraints 
can thus be removed to leading order.
 
Figure 2: The low energy contribution of \S IV.A. 
Diagram L1 represents the shift arising from treating photon emission below the cutoff
$\tilde{\lam}$ in second-order BSPT, where the intermediate electron-proton are bound by the
Coulomb potential.
Diagram L2 is an effective self-energy interaction (plus counterterm), arising from
the removal of photon emission above the cutoff $\tilde{\lam}$, treated
in first-order BSPT. $\overline{\beta}(2,l)$ is the average excitation energy of the $n=2$
state; see Eqs.~(\ref{eq:107}) and (\ref{eq:108}) and the discussion above them for details.

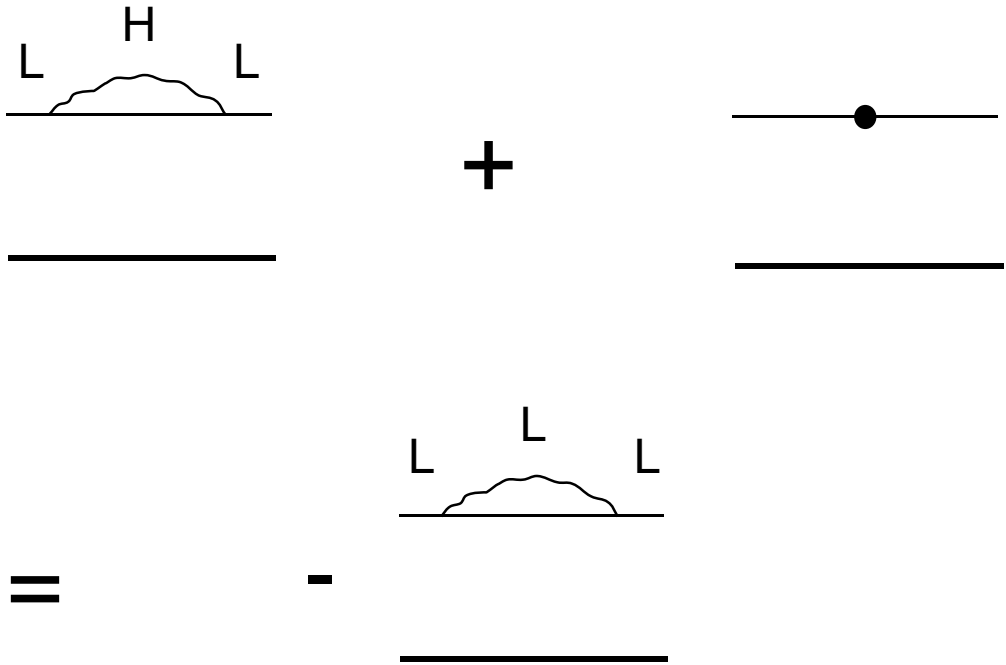
Figure 3: The sum of an effective self-energy
interaction arising from the removal of photon emission above 
the cutoff $\tilde{\lam}$
 and a counterterm. The counterterm is fixed by coupling coherence. The result
is the interaction in diagram L2 of Fig.~2.

Figure 4: The high energy contribution of \S IV.B. These are
third and fourth order effective interactions treated in first-order BSPT.
These effective interactions arise from the removal of photon emission above the cutoff
$\tilde{\lam}$.
`$b$' is an arbitrary scale, required to satisfy 
$m \alpha \ll b \ll m$, that was introduced to simplify the calculation. Note the 
$b$-independence of the result.
The total contribution is a sum of Fig.~2 and Fig.~4. Note the $\tilde{\lam}$-independence
of the combined result.

Figure 5: The $n=2$ hydrogen spectrum: fine-structure, Lamb-shift and hyperfine structure.
${\bf F}={\bf L}+{\bf S_e}+{\bf S_p}$.

\end{document}